\documentclass[%
 aip,
 amsmath,amssymb,
secnumarabic,%
 reprint,%
]{revtex4-1}

\usepackage[usenames, dvipsnames]{color}
\definecolor{myyellow}{RGB}{255, 255, 192}
\definecolor{mygreen}{RGB}{192, 255, 192}
\definecolor{myblue}{RGB}{192, 192, 255}

\usepackage{gensymb}
\usepackage{colortbl}
\usepackage{amsmath,amssymb,color}
\usepackage{mathrsfs}
\usepackage{upgreek}
\usepackage{dcolumn}
\usepackage{graphicx}
\usepackage{docs}%
\usepackage{bm}%
\usepackage[colorlinks=true,linkcolor=blue]{hyperref}%
\usepackage[usenames,dvipsnames]{xcolor}
\usepackage{float}
\usepackage[utf8]{inputenc}
\usepackage[T1]{fontenc}
\usepackage{mathptmx}
\usepackage{etoolbox}
\usepackage{placeins}
\usepackage{xcolor}

\usepackage{soul}

\expandafter\ifx\csname package@font\endcsname\relax\else
 \expandafter\expandafter
 \expandafter\usepackage
 \expandafter\expandafter
 \expandafter{\csname package@font\endcsname}%
\fi
\providecommand{\abs}[1]{\lvert#1\rvert}

\renewcommand{\epsilon}{\text{\usefont{OML}{cmr}{m}{n}\symbol{15}}}

\makeatletter
\def\@email#1#2{%
 \endgroup
 \patchcmd{\titleblock@produce}
  {\frontmatter@RRAPformat}
  {\frontmatter@RRAPformat{\produce@RRAP{*#1\href{mailto:#2}{#2}}}\frontmatter@RRAPformat}
  {}{}
}%
\makeatother
\begin{document}

\preprint{AIP/123-QED}


\title{Extensional rheometry of mobile fluids. Part II: Comparison between the uniaxial, planar and biaxial extensional rheology of dilute polymer solutions using numerically-optimized stagnation point microfluidic devices}
\author{Simon J. Haward}
 \email{simon.haward@oist.jp.}
 \affiliation{Okinawa Institute of Science and Technology, Onna, Okinawa 904-0495, Japan.}

\author{Stylianos Varchanis}
\affiliation{Okinawa Institute of Science and Technology, Onna, Okinawa 904-0495, Japan.}%

\author{Gareth H. McKinley}
\affiliation{Hatsopoulos Microfluids Laboratory, Department of Mechanical Engineering, Massachusetts Institute of Technology, 77 Massachusetts Avenue, Cambridge, MA 02139, USA}%

\author{Manuel A. Alves}
\affiliation{Departamento de Engenharia Qu\'{i}mica, ALiCE, CEFT, Faculdade de Engenharia da Universidade do Porto, Rua Dr. Roberto Frias, 4200-465 Porto, Portugal.}%

\author{Amy Q. Shen}
\affiliation{Okinawa Institute of Science and Technology, Onna, Okinawa 904-0495, Japan.}%

\date{7 April 2022}%
\revised{\today}%

\begin{abstract}
In Part I of this paper [Haward et al. submitted (2023)], we presented a new three-dimensional microfluidic device (the optimized uniaxial and biaxial extensional rheometer, OUBER) for generating near-homogeneous uniaxial and biaxial elongational flows. In this Part II of the paper, we employ the OUBER device to examine the uniaxial and biaxial extensional rheology of some model dilute polymer solutions. We also compare the results with measurements made under planar extension in the optimized-shape cross-slot extensional rheometer [or OSCER, Haward et al. Phys. Rev. Lett. (2012)]. In each case (uniaxial, planar and biaxial extension), we use micro-particle image velocimetry to measure the extension rate as a function of the imposed flow rate, and we measure the excess pressure drop across each device in order to estimate the tensile stress difference generated in the fluid. We present a new analysis, based on solving the macroscopic power balance for flow through each device, to refine the estimate of the tensile stress difference obtained from the measured pressure drop. Based on this analysis, we find that for our most dilute polymer sample, which is “ultradilute”, the extensional viscosity is well described by the finitely extensible non-linear elastic dumbbell model. In this limit, the biaxial extensional viscosity at high Weissenberg numbers (Wi) is half that of the uniaxial and planar extensional viscosities. At higher polymer concentrations, the experimental measurements deviate from the model predictions, which is attributed to the onset of intermolecular interactions as polymers unravel in the extensional flows. Of practical significance (and fundamental interest), elastic instability occurs at a significantly lower Wi in uniaxial extensional flow than in either biaxial or planar extensional flow, limiting the utility of this flow type for extensional viscosity measurement.
\end{abstract}

\maketitle

\section{\label{intro}Introduction}

For an incompressible Newtonian fluid of shear viscosity $\eta$, it is well known that the uniaxial extensional viscosity is $\eta_E=3 \eta$, the planar extensional viscosity is $\eta_P = 4 \eta$, and the biaxial extensional viscosity is $\eta_B = 6 \eta$, where the coefficients, 3, 4, and 6, are commonly referred to as the respective Trouton ratio $\text{Tr}$.~\cite{Trouton1906,Petrie2006} By contrast, for viscoelastic fluids such as polymer solutions and melts, these limiting values of the extensional viscosity are only approached at small rates of strain. At higher rates of strain, such that the dimensionless Weissenberg number $\text{Wi}\gtrsim 0.5$, the unraveling and orientation of polymer chains \cite{DeGennes1974,Hinch1974,Keller1985,Larson1989,Perkins1997} results in an increased elastic tensile stress difference $\Delta\sigma$ in the fluid and hence a non-linear increase in the extensional viscosity and apparent Trouton ratio $\text{Tr}_{app}$. 

Understanding how the extensional viscosity and $\text{Tr}_{app}$ for viscoelastic fluids at $\text{Wi}  > 0.5$ depends on the imposed mode of extension has interested a number of researchers over many years.~\cite{Stevenson1975,Meissner1982,Petrie1984,Demarmels1985,Jones1987,Khan1987,Khan1987b,Petrie1990,Isaki1991,Wagner1998,Nishioka2000,Kwan2001,Hachmann2003,Stadler2007,Shogin2021} A large number of studies have involved constitutive modeling, while most of the experimental work for the validation of those models has involved the extensional flow of polymer melts. For such highly elastic fluids, various instrumentation has been developed based on, e.g., the stretching of filaments or sheets of material held in rotary clamps,\cite{Meissner1981,Meissner1987,Hachmann2003} or by lubricated squeezing,\cite{Chatraei1981,Khan1987b,Nishioka2000} and the high elastic stresses resulting from the imposed deformation are quite readily measurable. By contrast, for less viscous, more mobile, viscoelastic fluids such as polymeric solutions, which can not be fixed in clamps and which generate relatively weak elastic stresses, the development of extensional rheometers is far more challenging.~\cite{James1993,Macosko1994,Haward2016} In this case, experimental comparisons between the response of viscoelastic fluids under different modes of extension are extremely rare.~\cite{Jones1987}

Extensional flows are potential flows characterized by diagonal rate-of-strain tensors $\text{\bf{D}}$, and come in three fundamental types. Uniaxial extension has one positive extensional axis and is compressional along the remaining directions, e.g.:

\begin{equation}
\text{\bf{D}}_U = 
\begin{pmatrix}
-\dot\varepsilon /2 & 0 & 0 \\
 0  & -\dot\varepsilon /2 & 0 \\
0 & 0 & \dot\varepsilon
\end{pmatrix}
.
\label{uni}
\end{equation}

\pagebreak
Planar extension has one neutral direction with equal and opposite extension and compression along two perpendicular directions, e.g.:

\begin{equation}
\text{\bf{D}}_P = 
\begin{pmatrix}
\dot\varepsilon  & 0 & 0 \\
 0  & -\dot\varepsilon  & 0 \\
0 & 0 & 0
\end{pmatrix}
.
\label{pla}
\end{equation}

Finally, biaxial extension is the kinematic reverse of uniaxial extension, having one compressional axis and with extension along the remaining directions, thus:

\begin{equation}
\text{\bf{D}}_B = 
\begin{pmatrix}
\dot\varepsilon_B & 0 & 0 \\
 0  & \dot\varepsilon_B  & 0 \\
0 & 0 & -2\dot\varepsilon_B
\end{pmatrix}
.
\label{bi}
\end{equation}

Note that here we use similar definitions for uniaxial and biaxial extension as those used by Meissner and coworkers,~\cite{Meissner1982,Demarmels1985} and adopted by Petrie,~\cite{Petrie1984,Petrie1990,Petrie2006} in the sense that we always consider the relevant strain rate metric on which material functions will be defined as that along the stretching direction(s). In accordance with Society of Rheology notation,~\cite{Dealy1984,Dealy1995} we place the subscript ``$B$'' on $\dot\varepsilon$ for biaxial extension. This serves to distinguish these expressions from alternative definitions, such as those suggested by Stevenson et al,~\cite{Stevenson1975} and by Bird,~\cite{Bird} for which biaxial extension is considered equivalent to uniaxial compression and is thus described by Eq.~\ref{uni} with a reversed sign, resulting in a strain rate of $\dot\varepsilon/2$ in the two orthogonal stretching directions.

The extra stresses that arise in viscoelastic polymer solutions for $\text{Wi}=\lambda \dot\varepsilon > 0.5$ (or $\text{Wi}= \lambda \dot\varepsilon_B > 0.5$ in biaxial flow) result from the  entropic elasticity of the polymer chains (with characteristic relaxation time $\lambda$), causing them to resist deformation and stretching. The hydrodynamically-forced stretching causes optical anisotropy in the fluid, often visible in experiments as flow-induced birefringence.~\cite{Fuller,Odell2007} Birefringence is thus an optical signature of the elastic stress in the fluid; indeed the two may be directly proportional in cases for which the stress-optical rule is obeyed.~\cite{Fuller} In hyperbolic stagnation point extensional flows (such as those described by Eqs.~\ref{uni} to \ref{bi}), due to the long residence time (or equivalently the large accumulated strain) available for polymers to unravel, the birefringence is predominantly aligned along the axes of positive extension rate. Here, the fluid has passed through (or near) the hyperbolic point at the coordinate origin, where the residence time in the straining flow (and hence the strain) is theoretically infinite. The localization of the birefringence about the stretching axes gives rise to the common descriptor of ``birefringent strand''.~\cite{Crowley1976,Keller1985,Harlen1990,Harlen1992,Remmelgas1999,Becherer2008,Becherer2009,Haward2012c,Haward2019b} It is the growth of the elastic stress along the stretching axes only that leads us to consider the relevant strain rate for determination of the extensional viscosity as also being that directed along the same axes.

\begin{figure}[!ht]
\begin{center}
\includegraphics[scale=0.4]{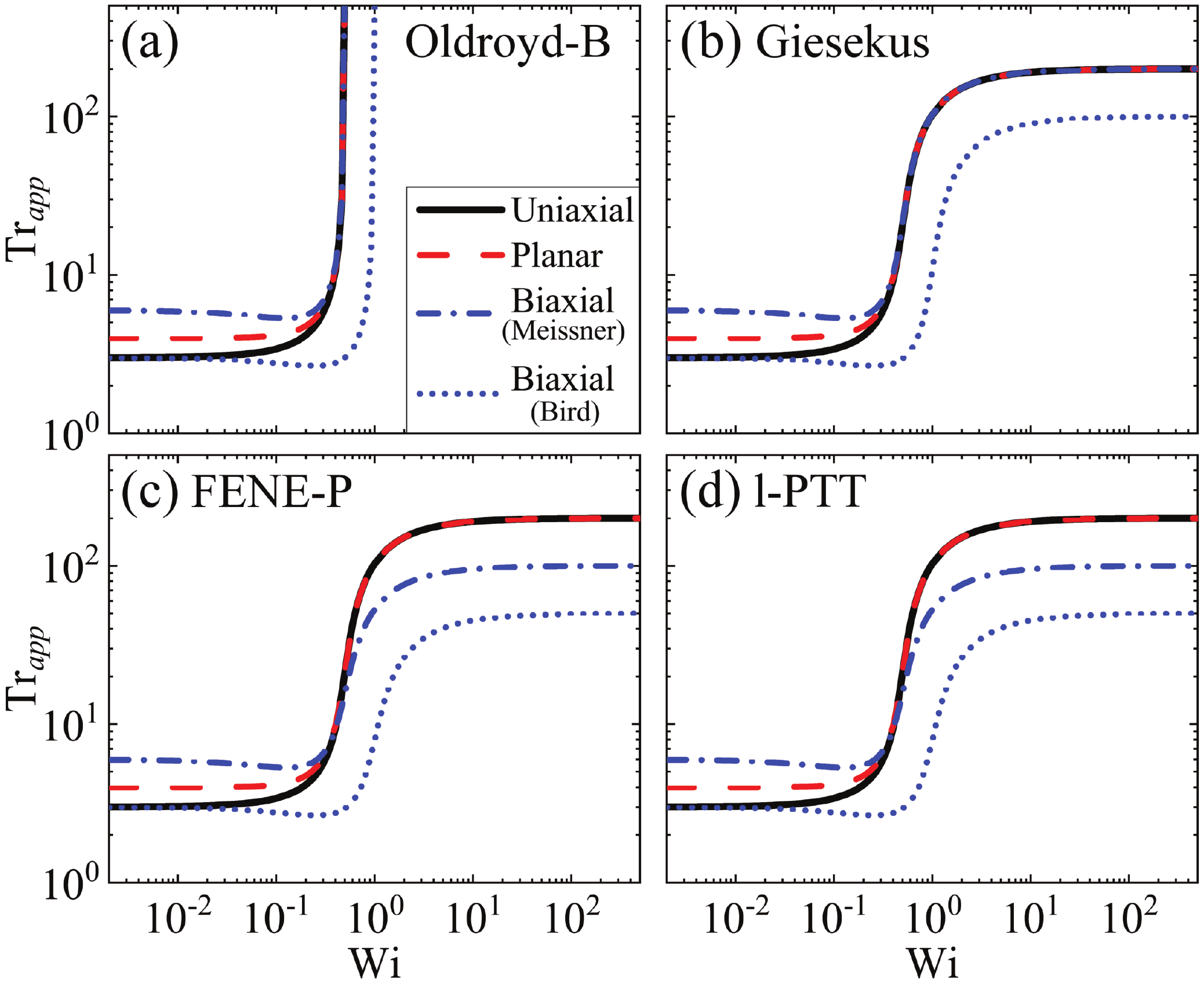}
\caption {The apparent Trouton ratio $\text{Tr}_{app}$ predicted in the three fundamental modes of extensional deformation by various common constitutive models used to describe dilute polymer solutions: (a) the Oldroyd-B model, (b) the Giesekus model (mobility factor $\alpha=0.01$), (c) the FENE-P model (extensibility $L=10$), and (c) the linear PTT (l-PTT) model (PTT extensibility parameter $\epsilon=0.01$). In all cases, the solvent viscosity is set to $\eta_s = 0$.
}
\label{predictions}
\end{center}
\end{figure}

To obtain a flow curve of extensional viscosity as a function of the strain rate, the extensional kinematics applied to the fluid should be both homogeneous in space and constant in time (i.e., \emph{persistent}).~\cite{Petrie2006} Thus, at each imposed extension rate, the polymer chains have sufficient time to achieve an equilibrium degree of stretching in the flow, and for the elastic stresses to reach a steady state as the strain $\varepsilon = \dot\varepsilon t \rightarrow \infty$ (or $\varepsilon_B = \dot\varepsilon_B t \rightarrow \infty$) over a residence time, $t \gg \lambda$. If the steady-state diagonal stress tensor resulting from the homogeneous extensional deformation is:

\begin{equation}
\pmb{\upsigma} = 
\begin{pmatrix}
\sigma_{xx} & 0 & 0 \\
 0  & \sigma_{yy}  & 0 \\
0 & 0 & \sigma_{zz}
\end{pmatrix}
,
\end{equation}
then the uniaxial extensional viscosity will be $\eta_{E}(\dot\varepsilon)~=~(\sigma_{zz}~-~\sigma_{xx})/\dot\varepsilon$, the planar extensional viscosity will be $\eta_{P}(\dot\varepsilon)~=
~(\sigma_{xx}~-~\sigma_{yy})/\dot\varepsilon$, and the biaxial extensional viscosity will be $\eta_{B}(\dot\varepsilon_B)~=
~(\sigma_{xx}~-~\sigma_{zz})/\dot\varepsilon_B$. The apparent Trouton ratio can be defined as $\text{Tr}_{app} = \eta_E/\eta_0$, $\eta_P/\eta_0$, or $\eta_B/\eta_0$ for uniaxial, planar, or biaxial extension (respectively), where $\eta_0$ is the steady shear viscosity of the fluid at zero shear rate.

Petrie (1990) obtained asymptotic results for the uniaxial and planar extensional viscosities given by various viscoelastic constitutive models commonly used to describe polymeric solutions.~\cite{Petrie1990} For all models examined, including the Phan-Thien and Tanner (PTT), the Giesekus, and the finitely extensible non-linear elastic dumbbell with Peterlin closure (FENE-P) model, the two extensional viscosities were equal at high $\text{Wi}$ (apart from the relatively small differences due to the different contribution of the solvent in each flow type). In fact, all viscoelastic constitutive models predict that at low deformation rates in uniaxial, planar, and biaxial elongation $\text{Tr}_{app}$ approaches the Newtonian limit of 3, 4, or 6 (respectively) as $\text{Wi} \rightarrow 0$, and that in all three flows, as the Weissenberg number exceeds 0.5, $\text{Tr}_{app}$ undergoes an abrupt increase (see Fig.~\ref{predictions}). For the infinitely extensible Oldroyd-B model (Fig.~\ref{predictions}(a)), $\text{Tr}_{app} \rightarrow \infty$ for $\text{Wi}\geq0.5$, in all cases. Models with a bounded elasticity show that under uniaxial and planar elongation,  $\text{Tr}_{app}$ approaches the same limiting plateau value as $\text{Wi} \rightarrow \infty$ (Fig.~\ref{predictions}(b-d)). However, there is a disagreement between the predictions of different models in terms of the high $\text{Wi}$ limit of $\text{Tr}_{app}$ in biaxial elongation. Some models, such as the Giesekus model, predict that $\text{Tr}_{app}$ will tend to the same high $\text{Wi}$ plateau in biaxial extension as it does in uniaxial and planar extension (Fig.~\ref{predictions}(b)). However, the FENE-P and PTT models predict that in biaxial extension the limiting value of $\text{Tr}_{app}$ at high $\text{Wi}$ will be one-half of that for uniaxial and planar extension (Fig.~\ref{predictions}(c,d)). Note that, using the alternative definition of the deformation rate tensor formulation for biaxial extension outlined by Bird \cite{Bird} results in a doubling of the Weissenberg number and a halving of the apparent Trouton ratio compared to the formulation of Meissner et al,~\cite{Meissner1982} as indicated by the dotted blue lines in Fig.~\ref{predictions}.~\cite{Shogin2021}

Due to the great difficulty associated with experimental extensional rheometry of low viscosity, mobile viscoelastic fluids such as dilute polymer solutions,~\cite{James1993,Macosko1994,Haward2016} these theoretical predictions are largely untested experimentally. Using a ``spin-line'' rheometer and a converging channel rheometer to generate uniaxial and planar extension, respectively, Jones et al (1987) found a ``satisfactory'' (meaning order-of-magnitude) correspondence between $\eta_{E}$ and $\eta_{P}$ for a variety of polymer solutions. \cite{Williams1985,Jones1987} However, the two measurement methods employed differed greatly in terms of how the deformation was applied, its spatial homogeneity, the range of deformation rates probed, and how the tensile stress was estimated. \cite{Williams1985,Jones1987} The authors themselves expressed apparent surprise at the agreement they obtained given the inherent problems with making such measurements, and remarking that they were not ``comparing like with like'' since one method was planar and the other uniaxial. To this day, a systematic experimental comparison of the uniaxial, planar and biaxial extensional responses of dilute polymer solutions using comparable measurement methods is still missing from the literature.

\begin{figure}[!t]
\begin{center}
\includegraphics[scale=0.55]{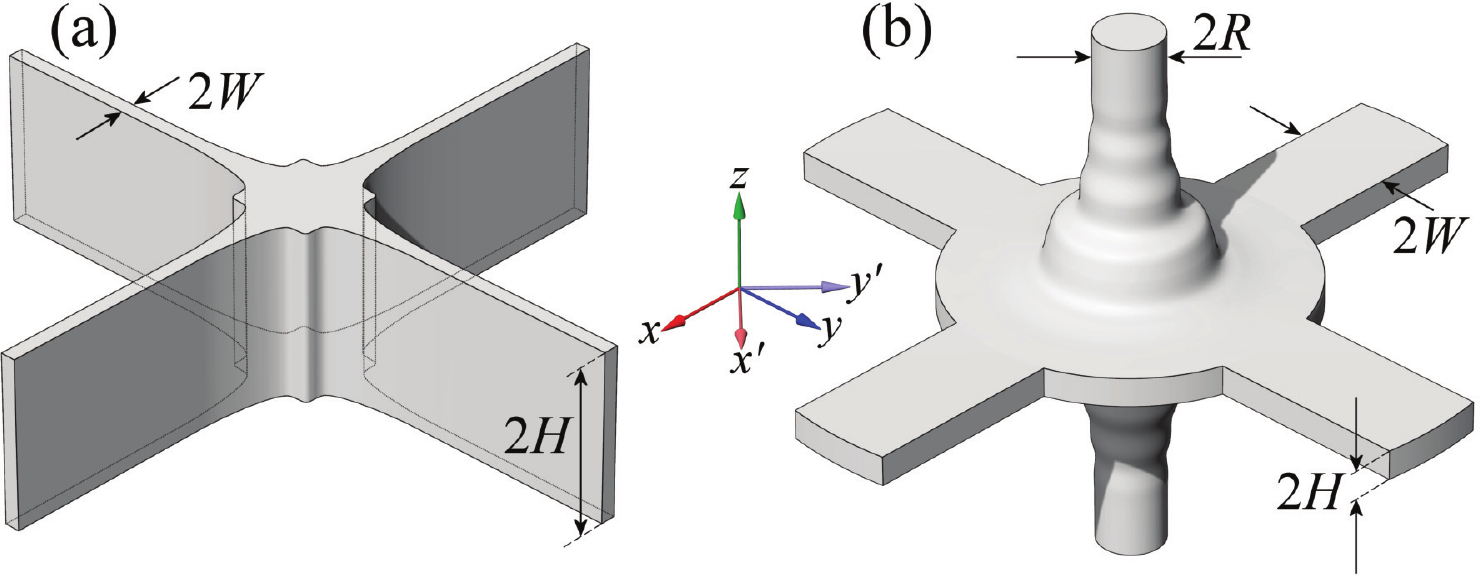}
\caption {Schematic illustrations of numerically optimized stagnation point elongational flow devices. (a) The optimized shape cross-slot extensional rheometer (OSCER) geometry consisting of two pairs of opposed planar inlet (outlet) channels (height $2H$, width $2W$) oriented along the $y$ ($x$) axes and joined by a numerically-determined profile designed to generate an optimal approximation to planar elongational flow. (b) The optimized uniaxial and biaxial extensional rheometer (OUBER), consisting of two pairs of opposed planar inlet (outlet) channels (height $2H$, width $2W$) oriented along the $x$ and $y$ axes and connected via a numerically-determined profile to a pair of opposing outlet (inlet) channels of circular cross-section oriented along the $z$ axis. Depending on the choice of imposed flow direction, the geometry can produce an optimal approximation to either uniaxial or biaxial elongational flow. For the OUBER geometry, along with the standard $(x,y,z)$ coordinate system, a $45^{\circ}$-rotated coordinate system $(x',y',z)$ is employed (see main text for details). The coordinate origin is located at the center of each device
}
\label{schematic}
\end{center}
\end{figure}

In the present work, we employ numerically optimized stagnation point microfluidic devices to make measurements of the uniaxial, planar and biaxial extensional viscosities of a variety of model solutions formulated from dilute concentrations of linear polymers. For planar extensional viscosity measurements we utilize the two-dimensional (2D) Optimized-Shape Cross-slot Extensional Rheometer (OSCER, Fig.~\ref{schematic}(a)),~\cite{Alves2008} which over the last decade has proven useful for characterizing the extensional rheology and flow behavior of a variety of viscoelastic fluids. \cite{Haward2012c,Haward2013b,Haward2016c} For uniaxial and biaxial extensional viscosity measurements we utilize the three-dimensional (3D) Optimized Uniaxial and Biaxial Extensional Rheometer (OUBER, Fig.~\ref{schematic}(b)) presented in Part I of this paper.~\cite{Haward2023} The devices are designed to provide close approximations to the respective deformation rate tensors (given in Eqs.~\ref{uni}-\ref{bi}) over multiple characteristic device lengthscales in each spatial direction. Both devices allow the strain rate to be controlled by simply varying the volumetric flow rate. They also both generate stagnation points at the center of the flow field, such that strain can accumulate indefinitely at the set strain rate (a requirement for measuring the extensional viscosity). Under all three modes of elongation, we employ micro-particle image velocimetry ($\upmu$-PIV) to confirm and quantify the extensional strain rates, coupled with pressure drop measurements designed to enable estimation of the elastic tensile stress difference. The comparable (microfluidic) size scales of the two devices allow similar extension rates to be obtained in each mode of extension while always keeping inertia negligible. 

We remark that experimental extensional viscosity measurements are always an approximation, and that extensional ``rheometers'' must always be considered ``indexers'' to some extent. In this work, for the first time we have assembled a pair of highly comparable and sophisticated indexers that permit a fair comparison between the extensional rheology of viscoelastic fluids under each of the three fundamental modes of extension. For the most dilute polymer solution that we test (which can be considered ``ultradilute''~\cite{Clasen2006}), our results at high $\text{Wi}>0.5$ indicate that $\eta_{E} \approx \eta_{P} \approx 2\eta_{B} $, in agreement with the prediction of the FENE-P constitutive model.  Of some interest, we observe that these elongational flows lose stability at different $\text{Wi}$ in each of the three flows (lowest in uniaxial and highest in biaxial extension), which we discuss in terms of the region occupied by the elastic ``birefringent strand'' that forms along the stretching axis (or over the stretching plane). These observations have important implications for viscoelastic constitutive modeling as well as for experimental extensional rheometry.

\section{Experimental Methods}
\label{ExpMeth}

\subsection{Microfluidic geometries}
\label{geom}

\subsubsection{Planar extensional flow OSCER device}

The OSCER device, shown schematically in Fig.~\ref{schematic}(a), has been described in detail in several prior works. \cite{Haward2012c,Haward2013b,Haward2016c} Briefly, the channel is cut in stainless steel by wire-electrical discharge machining and sealed about the $z$ direction with soda glass viewing windows. The channel has a uniform half-height $H=1$~mm and a characteristic half-width $W=0.1$~mm upstream and downstream of the optimized region. The channel shape is optimized over a region spanning $\abs{x},\abs{y} \leq 15W$, and generates a close approximation to pure planar elongation over a large portion of that region.\cite{Haward2012c,Haward2016c} The high aspect ratio of the device ($H/W=10$) gives a good approximation to a two-dimensional (2D) flow ensuring that the flow field is also uniform through most of the channel height.

\subsubsection{Uni- and biaxial extensional flow OUBER device}

The fabrication of an OUBER device (Fig.~\ref{schematic}(b)),~\cite{Haward2023} is achieved by the technique of selective laser-induced etching (SLE) in fused silica glass, \cite{Gottmann2012,Meineke2016,Burshtein2019} and is described in detail in Part I of this paper.~\cite{Haward2023}

The circular cross-section channels aligned along $z$ have a radius $R=0.4$~mm, while the four planar channels aligned along $x$ and $y$ each have half-width $W=0.64$~mm and half-height $H=0.16$~mm. The channel shape is optimized to provide almost uniform velocity gradients over a region spanning $\abs{x},\abs{y},\abs{z} \leq 5R$, and (depending on how the flow is imposed) generates a close approximation to either pure uniaxial or pure biaxial elongation over a large portion of that region.~\cite{Haward2023}

Note that, as depicted in Fig.~\ref{schematic}, it is natural to align the $x$ and $y$ axes with adjacent planar inlet/outlet channels. However, obtaining an experimental view inside of the OUBER device along either of those two directions is problematic with our current design. As described in Part I of the paper, optical access to the stagnation point region inside the device is only possible by viewing at $45^{\circ}$ to the $x$-axis.~\cite{Haward2023} Therefore, in our experimental setup we consider a coordinate system described by $(x',y',z)$, where $x' = \frac{1}{\sqrt{2}}(x+y)$, and $y' = \frac{1}{\sqrt{2}}(y-x)$ (also shown in Fig.~\ref{schematic}). 

\begin{figure}[!t]
\begin{center}
\includegraphics[scale=0.7]{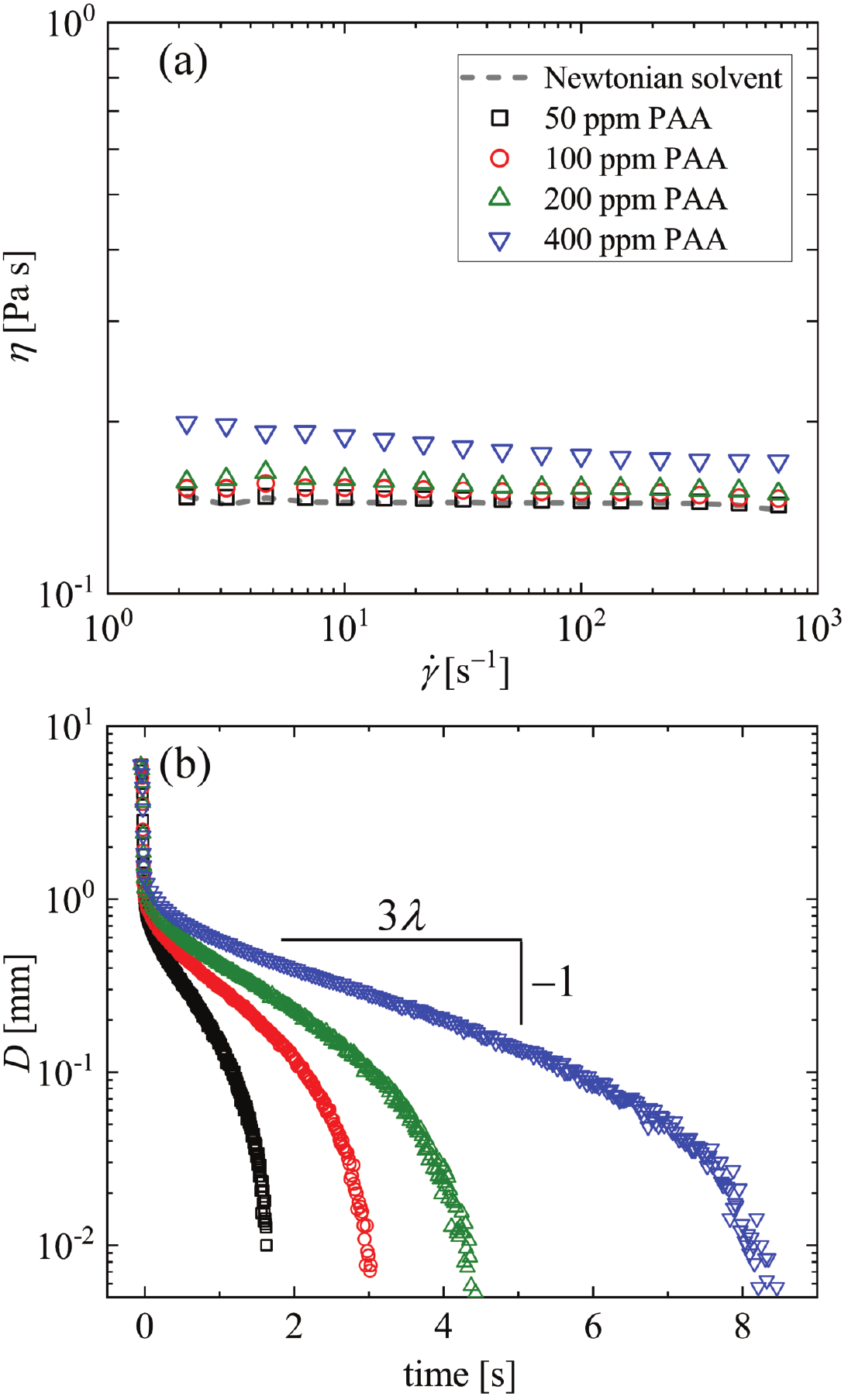}
\caption {Rheological response of the various test fluids employed. (a) Shear viscosity $\eta$ as a function of the applied shear rate $\dot\gamma$ of the Newtonian solvent (89.6\% glycerol in water, dashed line) and poly(acrylamide) (PAA) solutions at various polymer concentrations measured in steady shear using a stress-controlled TA Instruments DHR3 rotational rheometer with 40~mm diameter 1$^{\circ}$ cone-and-plate fixture. (b) Decay of the filament diameter $D$ as a function of time for the polymer solutions during capillary thinning in a CaBER device, used to obtain the extensional relaxation times $\lambda$ of the samples.
 } 
\label{flowcurves}
\end{center}
\end{figure}

\subsection{Test fluids}
\label{fluids}

Due to the surface curvature of the three-dimensional (3D) OUBER device, see Fig.~\ref{schematic}(b), clear imaging inside of the device (e.g., for performing flow velocimetry, as described below) requires that the channel be filled with a fluid of similar refractive index $RI$ as the fused silica glass ($RI=1.4584$).~\cite{Malitson1965} A mixture of 89.6~wt\% glycerol and 10.4~wt\% water, with $RI=1.4582$ at $25^{\circ}$C  (measured using an Anton-Paar Abbemat MW refractometer operating at 589~nm) is found to be a sufficiently close match. The 89.6\,:\,10.4~wt\% glycerol\,:\,water mixture (with density $\rho = 1231$~kg~m$^{-3}$ and viscosity $\eta_s = 0.143$~Pa~s, Fig.~\ref{flowcurves}(a)) is used as both a Newtonian reference fluid and also as a solvent for viscolelastic polymeric test solutions. 

The polymer sample used is a nonionic poly(acrylamide) (PAA) of molecular weight $M \approx 5$~MDa obtained from Sigma-Aldrich. Polymer solutions are prepared at four different concentrations $c = 50,~100,~200$, and 400 parts-per-million (ppm) by first dissolving the required mass of dry polymer powder in the aqueous component of the solvent, before adding the mass of glycerol necessary to achieve the final desired composition. To avoid mechanical degradation of the polymer during the solution preparation, mechanical stirring is not used. Rather the fluids are mixed by gentle agitation on a roller-mixer (Ika, Japan). Typically, 24~h is required for complete dissolution of the polymer powder into the water, and a further 24~h for complete mixing with the glycerol. Subsequent to preparation, the fluids are stored at $5^{\circ}$C in unlit conditions, and are discarded if not used within one month. 

The concentration regime and equilibrium conformation of the PAA in solution can be estimated based on the number of backbone bonds $n = 2 M/m \approx 140,000$ (where $m = 71$~Da is the monomer molecular weight), the average length per bond $l = 0.154$~nm, and the characteristic ratio $C_\infty=6.9$.~\cite{Scholtan1954,Winston1980,Kulicke1982} From this information, it is possible to estimate the contour length $L_c =nl \approx 21.6~\upmu$m and the equilibrium mean-square end-to-end length $\langle r_0^2\rangle=C_{\infty}nl^2\approx23,000~\text{nm}^2$. The radius of gyration is then $R_g=\frac{1}{\sqrt{6}}\langle r_0^2\rangle^{1/2} \approx 62$~nm, which can be used to estimate the overlap concentration $c^*=M/N_A(2R_g)^3\approx4400$~ppm (where $N_A$ is Avogadro's number).~\cite{Graessley1980} It can also be estimated that to achieve full stretch of the polymer chain, the end-to-end separation needs to be increased from its equilibrium value by an extensibility factor (or stretch ratio) of $L = L_c/\langle r_0^2\rangle^{1/2}\approx143$. We note that these molecular parameters are estimated based on the value of $C_\infty$ reported for PAA in water at $25^{\circ}$C.~\cite{Kulicke1982} However, recent molecular dynamics simulations indicate that the PAA chain adopts a roughly similar conformation in 90~wt\% aqueous glycerol as it does in pure water.~\cite{Hopkins_S_2020} Therefore, we have some confidence that our test fluids should be safely in the dilute solution regime with $0.011 \lesssim c/c^* \lesssim 0.088$, and that the PAA chains should be highly extensible. 

The steady shear rheology of the polymeric test fluids is measured using a stress-controlled DHR3 rotational rheometer (TA Instruments Inc.) fitted with a 40~mm diameter 1$^{\circ}$ angle cone-and-plate geometry (see Fig.~\ref{flowcurves}(a)). Over the range of accessible shear rates, the fluids each have a near-constant viscosity, close to that of the solvent. For this reason, we take the viscosity $\eta$ of each fluid as being the average of the respective data shown in Fig.~\ref{flowcurves}(a). The relaxation times $\lambda$ of the fluids are assessed by means of capillary thinning measurements using a CaBER device (Thermo-Haake). \cite{Anna2001b} The device is fitted with 6~mm diameter plates with the initial separation set to 1~mm and the final separation to 6~mm. Curves of the filament diameter at the midpoint between the plates $D$ as a function of time are shown in Fig.~\ref{flowcurves}(b). The value of $\lambda$ is extracted from the time constant of the exponential decay of the filament diameter observed within the elasto-capillary thinning regime. \cite{Anna2001b} The values of $\eta$ and $\lambda$ obtained for each polymeric fluid are given in Table~\ref{tab1}.

\subsection{Flow control and dimensionless groups}
\label{control}

The test fluids are driven through the microfluidic OSCER and OUBER devices using 29:1 gear ratio neMESYS low pressure syringe pumps (Cetoni, GmbH) to control the volumetric flow rate through each individual channel. For planar extensional flow in the OSCER device, two pumps are used to impose equal volumetric flow rates $Q$ into each of the two inlet channels, while two pumps withdraw fluid at equal and opposite rates from each of the two outlet channels. For uniaxial (biaxial) extensional flow in the OUBER device, two pumps are used to impose equal volumetric flow rates $Q$ through each of the two circular cross-section outlet (inlet) channels, while four pumps impose equal volumetric flow rates $Q/2$ through each of the four planar inlet (outlet) channels. The pumps are fitted with Hamilton Gastight syringes of appropriate volumes such that the specified ``pulsation free'' dosing rate is always exceeded. Connections between the syringes and the microfluidic devices are made using flexible Tygon tubing. 

\begin{table}
\caption{\label{tab1}Values of viscosity $\eta$, solvent-to-total viscosity ratio $\beta$, and relaxation time $\lambda$ obtained from rheological characterization of the PAA solutions. }
\begin{ruledtabular}
\begin{tabular}{c c c c c}
PAA concentration [ppm]   &   $c/c^*$ & $\eta$ [Pa~s] & $\beta = \eta_s/\eta$ &  $\lambda$ [s]      \\
\hline
50 & 0.011 & 0.146 & 0.98 & 0.22           \\
100 & 0.022  & 0.151 & 0.95 & 0.38      \\
200 &  0.044 & 0.155 & 0.92 & 0.54    \\
400 & 0.088 & 0.181 & 0.79 & 1.03     \\

\end{tabular}
\end{ruledtabular}
\end{table}

For an imposed volumetric flow rate $Q$ in each channel of the OSCER device, the average flow velocity is $U=Q/4WH$ and the expected (or nominal) extension rate based on a Newtonian flow field prediction is given by $\dot\varepsilon_{nom}=0.1U/W$.~\cite{Haward2012c,Haward2013b,Haward2016c}

For the OUBER device, we consider the characteristic average flow velocity $U$ as being that in the two channels of circular cross-section, so that for an imposed volumetric flow rate $Q$ in each of those channels, $U=Q/\uppi R^2$. The expected nominal extension rates are $\dot\varepsilon_{nom}=0.4U/R$ for uniaxial extension, and $\dot\varepsilon_{B,nom}=0.2U/R$ for biaxial extension.~\cite{Haward2023}

The Reynolds number $\text{Re}$ describes the relative strength of inertial to viscous forces in the flow experiments. In the OSCER device, we define $\text{Re}=\rho U D_h / \eta$, where $D_h=2WH/(W+H)$ is the hydraulic diameter of the rectangular channels. The maximum Reynolds number reached in experiments using the OSCER device is $\text{Re} \approx 0.1$. In the case of the OUBER device, we define $\text{Re} = 2 \rho U R / \eta$, and the maximum values reached are $\text{Re} \approx 0.2$ (uniaxial extension), and $\text{Re} \approx 0.7$ (biaxial extension). Since $\text{Re} < 1$ in all experiments, inertial effects in the flow are considered negligible.

The Weissenberg number describes the relative strength of elastic to viscous forces in the flow and can be quantified by the product of the extension rate and the relaxation time $\lambda$. However, since we only have \emph{a priori} knowledge of the nominal extension rate, it is convenient to first also define a nominal Weissenberg number as $\text{Wi}_{nom} = \lambda \dot\varepsilon_{nom}$ in uniaxial and planar elongation, and $\text{Wi}_{nom} = \lambda \dot\varepsilon_{B,nom}$ in biaxial elongation.

Typically in elongational flows, it is found that polymer stretching for $\text{Wi}_{nom} \gtrsim 0.5$ will modify the flow field compared to the Newtonian case, resulting in a reduction of the true extension rate below its nominal value.~\cite{Mackley1978,Dunlap1987,Remmelgas1999,Haward2012c,Haward2013b,Haward2016c} In the present work, micro-particle image velocimetry ($\upmu$-PIV) experiments (described in Sec.~\ref{PIV}) will be used to directly measure the true extension rate (or velocity gradient) along the stretching axis $\dot\varepsilon$ (or $\dot\varepsilon_B$), allowing the true Weissenberg number to be evaluated as $\text{Wi} = \lambda \dot\varepsilon$ in uniaxial and planar elongation, and $\text{Wi} = \lambda \dot\varepsilon_{B}$ in biaxial elongation.

\subsection{Microparticle image velocimetry}
\label{PIV}

Quantitative measurement of the flow field in each extensional flow configuration is achieved using microparticle image velocimetry ($\upmu$-PIV, TSI Inc., MN).~\cite{Wereley2005,Wereley2010} For this purpose, the test fluids are seeded with a low concentration ($c_p \approx 0.02$~wt\%) of $3.2~\upmu$m diameter red fluorescent tracer particles (Fluor-Max, Thermo Scientific) with excitation (emission) wavelength 542~nm (612~nm). The plane of interest within the geometry (i.e., the $xy$ midplane in the OSCER geometry, and the $y'=0$ plane in the OUBER geometry) is brought into focus on an inverted microscope (Nikon Eclipse Ti) with a $4 \times$ magnification, $\text{NA}=0.13$ numerical aperture Nikon PlanFluor objective lens. Under these conditions, the measurement depth over which microparticles contribute to the determination of the velocity field is $\delta_m \approx 180~\upmu$m. \cite{Meinhart2000} Excitation with a dual-pulsed Nd:YLF laser with a wavelength of 527~nm induces the emission of particle fluorescence, which is detected by a high speed camera (Phantom MIRO, Vision Research). The camera is operated in frame-straddling mode and is synchronized with the laser in order to acquire pairs of particle images corresponding to pairs of laser pulses separated by a small time $\Delta t$. The value of $\Delta t$ is varied inversely to the imposed flow rate and set so that the average displacement of particles between the two images in each pair is always $\approx4$~pixels. In this work we are only concerned with steady flows, so at each flow rate tested 50 image pairs are acquired and are processed using an ensemble average cross-correlation PIV algorithm (TSI Insight 4G) in order to reduce noise. A recursive Nyquist criterion is employed with a final interrogation area of $16 \times 16$ pixels to enhance the spatial resolution and obtain two components of the velocity vector $\bf{u}$ spaced on a square grid of $26.6~\upmu$m~$\times 26.6~\upmu$m. In the OSCER device, the obtained components of $\bf{u}$ are $u$ and $v$ (the $x$ and $y$ component, respectively). In the OUBER device, the obtained components of $\bf{u}$ are $u'$ and $w$ (the $x'$ and $z$ component, respectively). Subsequent to data acquisition, the software Tecplot Focus (Tecplot Inc., WA) is used for generation of velocity contour plots and streamline traces and for extraction of velocity profiles. 

\subsection{Pressure drop measurements and extensional rheometry}
\label{press}

Pressure drop measurements are made using a 35.5~kPa wet-wet differential pressure sensor (Omega Engineering Inc.) connected across one inlet and one outlet of each device. Pressure taps are taken by installing T-junction connectors in the upstream and downstream tubing connecting between the fluidic device and the syringes driving the flow. At each imposed flow rate in each extensional flow configuration (uniaxial, planar and biaxial), two independent measurements of the pressure drop are made. The first pressure drop measurement (labeled $\Delta P_{tot}$) is made with flow imposed in all the channels of the device (i.e., with the device in normal operation, as described in Sec~\ref{control}) and provides an estimate of the total stress. This combines stresses due to the shear induced by the walls of the channel and connecting tubing, as well as any extra stress due to the elongational kinematics in the flow. A second measurement (labeled $\Delta P_{sh}$) is made with half of the inlet channels and half of the outlet channels disabled and allows estimation of the shear stresses only. It is important to state that in the OUBER device, two adjacent (not opposing) planar channels are disabled during the measurement of $\Delta P_{sh}$ in order to avoid the formation of a `T-channel-like' flow configuration which would retain a stagnation point in the center of the device.

\begin{figure*}[!ht]
\begin{center}
\includegraphics[scale=0.8]{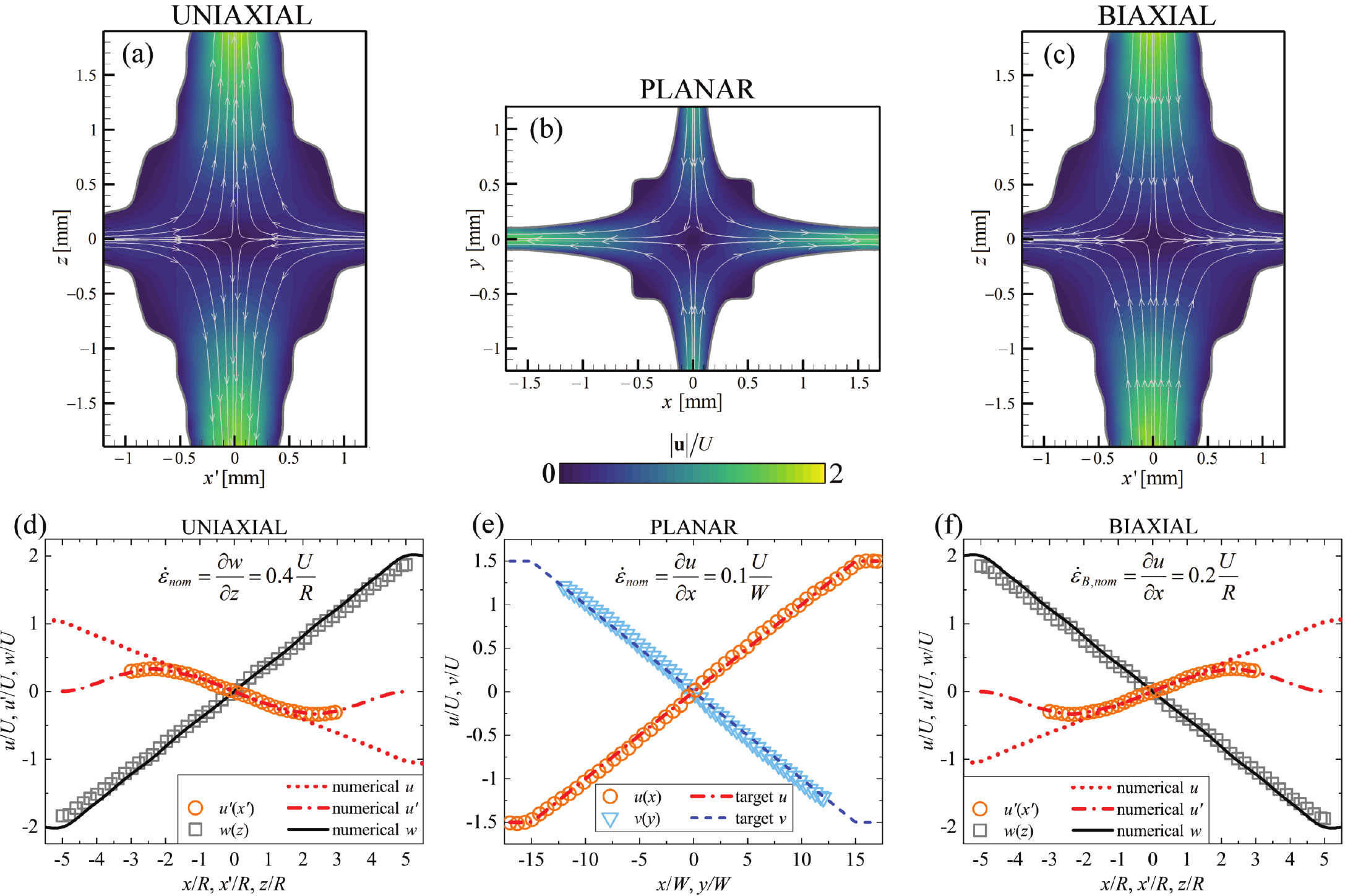}
\caption {Normalized velocity fields with superimposed streamlines for creeping flow ($\text{Re}<0.05$) of the Newtonian solvent in (a) uniaxial, (b) planar, and (c) biaxial extension. Parts (d), (e) and (f) show the respective normalized velocity profiles measured along the flow axes (data points), which compare favorably against the target numerical profiles (lines). The nominal elongation rate in each case is indicated within the respective plot.
} 
\label{NewtPIV}
\end{center}
\end{figure*}

From the raw pressure drop measurements, we obtain an excess pressure drop $\Delta P_{ex} = \Delta P_{tot} - \Delta P_{sh}$, which we assume arises predominantly due to the extensional kinematics present in the flow field during the measurement of $\Delta P_{tot}$. Of course, this differential measurement is not able to quantify each individual  component of the diagonal stress tensor in order to precisely evaluate the principal stress difference $\Delta \sigma$ required to compute the extensional viscosity (see Sec.~\ref{intro}). In previous works involving planar extensional flows in the OSCER device, and also in the standard cross-slot geometry, it has simply been assumed that $\Delta P_{ex} \approx \Delta \sigma$, thus the planar extensional viscosity has been computed as $\eta_{P} \approx \Delta P_{ex}/\dot\varepsilon$.~\cite{Haward2011,Haward2012a,Haward2012c,Haward2013b} Some support for this assumption has been shown using birefringent polymer solutions, for which direct proportionality has been shown between $\Delta P_{ex}$ and the birefringence $\Delta n$ measured at the stagnation point. The constants of proportionality approximately matched with the known stress-optical coefficients $C$, of the respective fluids, suggesting that $\Delta P_{ex} \approx \Delta n / C = \Delta \sigma$.~\cite{Sharma2015,Fuller}

In this work, we attempt to more properly relate $\Delta P_{ex}$ to $\Delta \sigma$ by considering the macroscopic power balance for flow through each of our geometries, thus enabling a more accurate estimation of the extensional viscosity to be obtained from the experimental pressure drop measurements (see details in Sec.~\ref{etaE}).

\section{Results}
\label{Res}

\subsection{Newtonian flow field characterization}

\begin{figure*}[!ht]
\begin{center}

\includegraphics[scale=0.8]{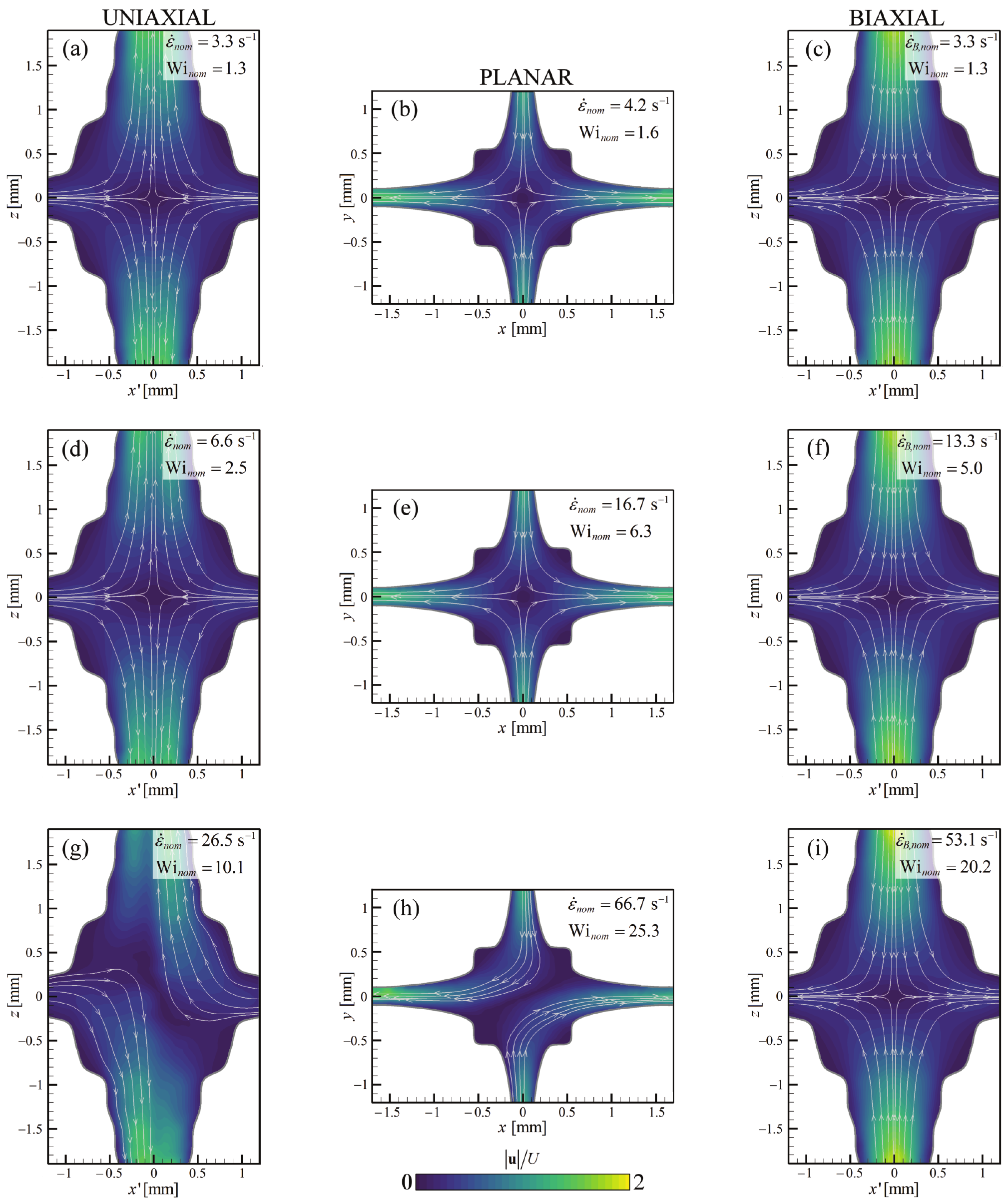}
\vspace{-0.1in}
\caption {Normalized velocity fields with superimposed streamlines for the flow of the 100~ppm poly(acrylamide) solution in (a,d,g) uniaxial, (b,e,h) planar, and (c,f,i) biaxial extension at the nominal extension rates and nominal Weissenberg numbers indicated. 
} 
\label{100ppmPIV}

\end{center}
\end{figure*}

For completeness, we commence the presentation of the experimental results by showing the Newtonian flow field in each extensional flow configuration (Fig.~\ref{NewtPIV}). Normalized fields of the velocity magnitude measured by $\upmu$-PIV, with streamlines superimposed to indicate the direction of flow, are shown in Fig.~\ref{NewtPIV}(a), (b) and (c) for uniaxial, planar and biaxial extension, respectively, at Reynolds numbers $\text{Re} < 0.05$. In each case, the velocity field is symmetric about the flow axes, with a stagnation point at the coordinate origin, as expected. Velocity profiles extracted along the flow axes are shown in Fig.~\ref{NewtPIV}(d), (e) and (f) (below the respective velocity field), in comparison with the numerical predictions for Newtonian creeping flow (available in Refs. \citenum{Alves2008} and \citenum{Haward2023}, for the OSCER and the OUBER, respectively). It is clear that over the measurable ranges of the accessible flow axes in each device configuration, the experimental velocity profiles agree very well with the respective numerical predictions, giving confidence that the microfluidic devices and the experimental setup are performing satisfactorily. It should be noted that over the $z=0$ plane in the OUBER device, the 2D $\upmu$-PIV measurement provides the $u'(x')$ velocity profile, which agrees with the $u(x)$ profile for $\abs{x'} \lesssim R$ and $\abs{x} \lesssim R$, as shown in Part I.~\cite{Haward2023} In other words, the velocity field has good axisymmetry over a radial distance of $r = \sqrt{x^2 + y^2} \approx R$ about the $z$-axis.

\subsection{Polymer solution flow field characterization}
\label{flowfield}

We proceed to examine the flow field in the case of the viscoelastic PAA-based test solutions. Fig.~\ref{100ppmPIV} shows normalized velocity magnitude fields obtained in each of the three extensional flow configurations for the 100~ppm PAA solution over a range of imposed nominal extension rates. At lower nominal rates, such that $\text{Wi}_{nom}$ is only slightly above unity, the flow field in each case (uniaxial, planar, and biaxial, shown in Fig.~\ref{100ppmPIV}(a), (b), and (c), respectively) appears to be rather similar to that observed for the flow of Newtonian fluid (shown in Fig.~\ref{NewtPIV}(a), (b), and (c), respectively). However, close inspection of the velocity magnitude contours for uniaxial extension of the polymer solution (Fig.~\ref{100ppmPIV}(a)) reveals a local minimum in the velocity along the stretching axis (i.e., along the $x'=0$ centerline). In contrast, for Newtonian flow, the Poiseuille-like velocity profile is maximal along the center of the outlet channels. This local minimum in velocity along the stretching direction is not evident in either the planar (Fig.~\ref{100ppmPIV}(b)) or biaxial (Fig.~\ref{100ppmPIV}(c)) flows at these rather low imposed values of $\text{Wi}_{nom}$. For the uniaxial flow, increasing the nominal Weissenberg number to $\text{Wi}_{nom}=2.5$ (Fig.~\ref{100ppmPIV}(d)), results in the centerline minimum of the velocity profile across the outlet channels becoming more pronounced. For the planar extensional flow at $\text{Wi}_{nom}=6.3$ (Fig.~\ref{100ppmPIV}(e)), and for biaxial extension at a similar $\text{Wi}_{nom}=5.0$ (Fig.~\ref{100ppmPIV}(f)), still no obvious difference from Newtonian flow can be discerned (see Fig.~\ref{NewtPIV}(b,c) for comparison). At sufficiently high $\text{Wi}_{nom}$, uniaxial and planar extensional flows of the 100~ppm PAA solution exhibit elastic instabilities manifested as an asymmetry of the flow. For uniaxial extension, this occurs for a critical nominal Weissenberg number $\text{Wi}_{nom,c} \approx 5$ as a distinct distortion of the streamlines close to the stagnation point, and develops with increasing $\text{Wi}_{nom}$ into the strong flow asymmetry shown in Fig.~\ref{100ppmPIV}(g) for $\text{Wi}_{nom}=10.1$. For planar extension, a somewhat higher critical value $\text{Wi}_{nom,c} \approx 13$ is necessary before the flow exhibits instability. Fig.~\ref{100ppmPIV}(h) shows a strongly asymmetric flow state observed in planar extension for $\text{Wi}_{nom}=25.4$. In contrast, for biaxial extension, no obvious sign of instabilty is observed even at the highest achievable nominal Weissenberg number $\text{Wi}_{nom}=20.2$ (Fig.~\ref{100ppmPIV}(i)); indeed, the kinetics appear to remain essentially Newtonian-like. 

\begin{figure*}[!ht]
\begin{center}
\includegraphics[scale=0.75]{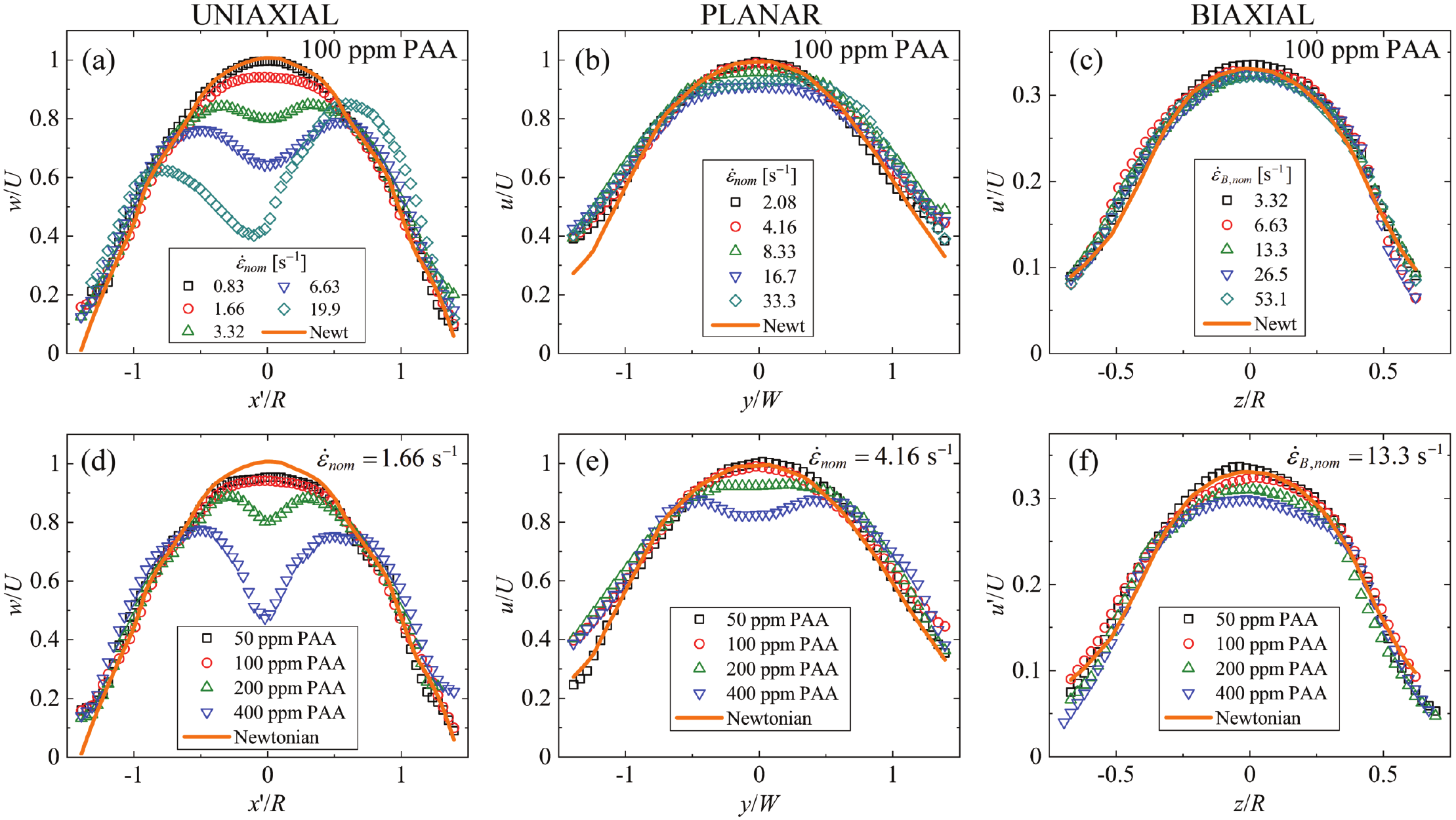}
\caption {Illustration of the flow modification along the stretching direction resulting from the flow of poly(acrylamide) solutions in (a,d) uniaxial, (b,e) planar, and (c,f) biaxial extension. Flow velocity profiles are measured across a device outlet 1~mm downstream of the stagnation point. (a,b,c) show normalized profiles of the streamwise flow velocity for the 100~ppm PAA solution at various nominal extension rates and compared against the result for the Newtonian solvent. (d,e,f) show normalized profiles of the streamwise flow velocity for all the tested polymer solutions at the highest nominal extension rate (indicated in the plot) for which all are deemed to be steady and symmetric, again compared against the result for the Newtonian solvent. 
} 
\label{PAA_trans_PIV}
\end{center}
\end{figure*}

The flow asymmetry observed in the OSCER device (Fig.~\ref{100ppmPIV}(h)) has been observed previously in various experiments involving planar stagnation point extensional flows.~\cite{Gardner1982,Arratia2006,Poole2007,Rocha2009,Haward2012a,Haward2013,Haward2016c} It is considered to be a purely elastic phenomenon driven by elastic tensile stress on the strongly curving streamlines that pass through the birefringent strand in the vicinity of the stagnation point, a mechanism consistent with the well-known elastic instability criterion introduced by McKinley and coworkers.~\cite{Pakdel1996,McKinley1996,Oztekin1997,Haward2016c} The asymmetric flow state observed under uniaxial extension in the OUBER device (Fig.~\ref{100ppmPIV}(g)) appears to be similar in form to that observed in the OSCER (Fig.~\ref{100ppmPIV}(h)), but it is unclear exactly how this asymmetry is oriented in the 3D `axisymmetric' flow field of the OUBER. A detailed investigation is beyond the scope of the present work and will require careful visualization in 3D, possibly using microtomographic flow velocimetry.~\cite{Haward2023} A cursory investigation indicates that the asymmetry in the OUBER device (Fig.~\ref{100ppmPIV}(g)) is steady in time and not rotating around the stretching axis. Most likely, it selects a favored orientation due to the presence of the four planar inlet channels which break the perfect axisymmetry of the flow away from the $z$-axis, similar to the instability reported by Afonso et al.~\cite{Afonso2010} in numerical simulations of viscoelastic flow in a 6-arm cross-slot.

In this work, which is focused on extensional rheometry, we wish to avoid elastic instabilities. Each viscoelastic test fluid is driven to the point of instability only in order to determine the limiting values of $\text{Wi}_{nom}$ up to which the extensional flow field generated around the stagnation point remains stable and symmetric. For experimental determination of the extensional viscosity under each extensional flow configuration (Sec.~\ref{etaE}), the range of extension rates is restricted to $\text{Wi}_{nom} < 0.5 \text{Wi}_{nom,c}$ in order to ensure the measurement is made while the flow is stable and symmetric.

\begin{figure*}[!ht]
\begin{center}
\includegraphics[scale=0.75]{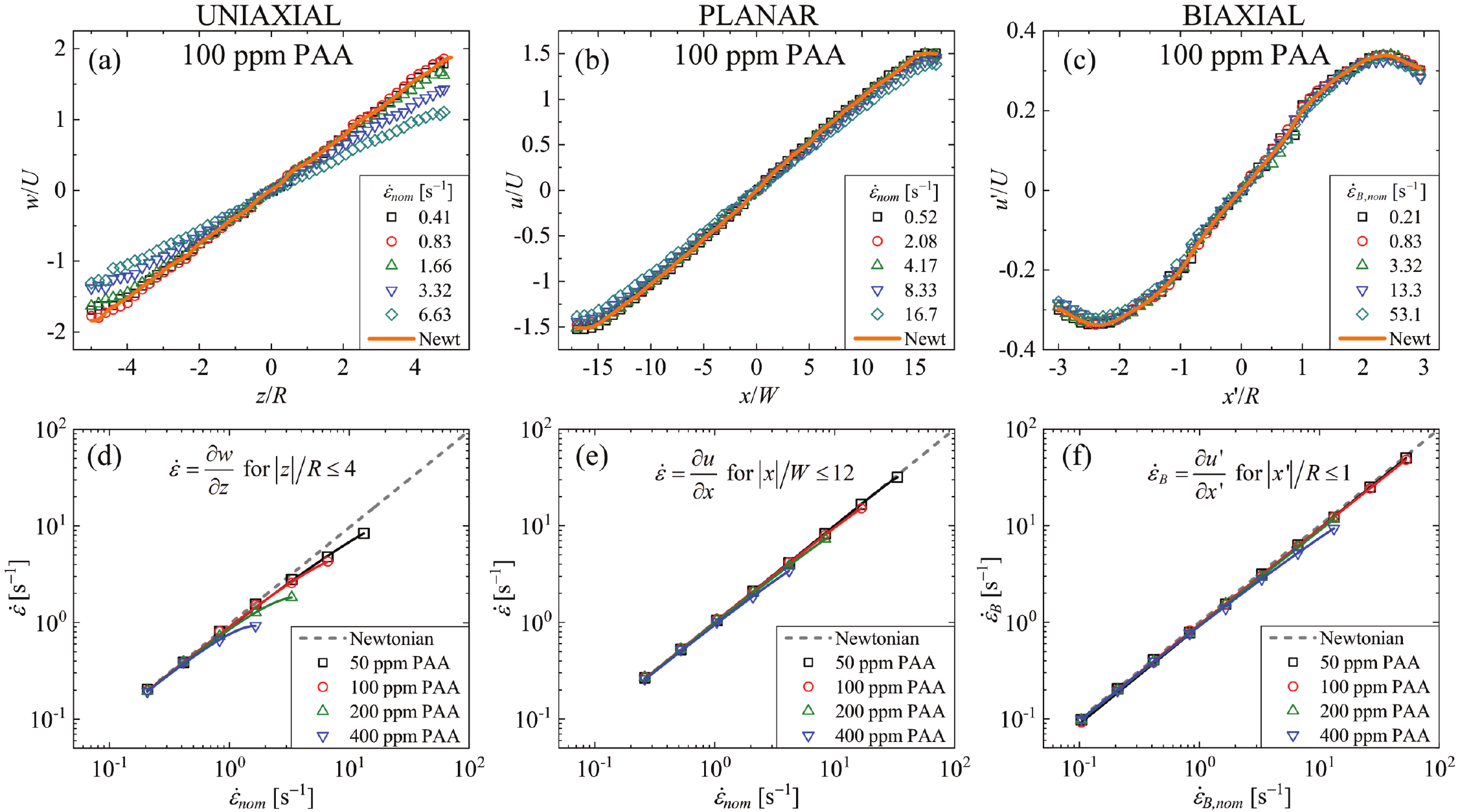}
\caption {Quantification of extension rates determined from velocity fields measured with poly(acrylamide) solutions in (a,d) uniaxial, (b,e) planar, and (c,f) biaxial extension. (a,b,c) show normalized streamwise velocity profiles along the stretching axes measured for the 100~ppm PAA solution at various nominal extension rates and compared against the result for the Newtonian solvent. (d,e,f) show the measured extension rates as a function of the average flow velocity for all the tested polymer solutions and compared against the result for the Newtonian solvent. The extension rate in each case is determined by averaging the velocity gradient on the relevant axis over the spatial domain indicated in the respective plot. Solid lines are fits to the experimental data points of the form described in the main text. 
} 
\label{PAA_PIV}
\end{center}
\vspace{-0.1in}
\end{figure*}

The modification of the Newtonian flow field by the presence of polymer (i.e., the development of local minima in the velocity profiles, mentioned above) is rendered more apparent by extracting velocity profiles across the channel outlets. Fig.~\ref{PAA_trans_PIV} shows normalized profiles of the streamwise flow velocity taken across a channel outlet 1~mm downstream of the stagnation point. Such profiles for the 100~ppm PAA solution (for which example velocity fields at different extension rates are provided in Fig.~\ref{100ppmPIV}) are shown in comparison with the Newtonian case for uniaxial, planar, and biaxial extension in Fig.~\ref{PAA_trans_PIV}(a), (b), and (c), respectively. During uniaxial extension in the OUBER device (Fig.~\ref{PAA_trans_PIV}(a)), the velocity profile obtained for the polymer solution agrees well with that of the Newtonian fluid at low $\dot\varepsilon_{nom}$, but an increasingly pronounced local minimum develops at $x'=0$ as the flow rate through the device is increased. At high $\dot\varepsilon_{nom}$, the profile becomes asymmetric about $x'=0$ as the symmetric base flow becomes unstable and the elastic asymmetry (illustrated in Fig.~\ref{100ppmPIV}(g)) develops. For planar extension in the OSCER device (Fig.~\ref{PAA_trans_PIV}(b)), the velocity profile across the channel outlet for the 100~ppm PAA solution again agrees well with that of the Newtonian fluid at low $\dot\varepsilon_{nom}$. In this case, with increasing $\dot\varepsilon_{nom}$, there is a progressive modification to the velocity profile, with some flattening of the central peak at $y=0$, but the profile remains almost parabolic. For the 100~ppm PAA solution in biaxial extension in the OUBER device, no significant difference is noticable compared to the Newtonian flow profile, even up to the highest values of $\dot\varepsilon_{B,nom}$ examined (Fig.~\ref{PAA_trans_PIV}(c)).

To illustrate the effects of increasing the polymer concentration, Fig.~\ref{PAA_trans_PIV}(d), (e), and (f) shows the velocity profiles across the channel outlets for all of the tested fluids in uniaxial, planar, and biaxial extension (respectively). In each case the profiles are shown for a fixed value of the nominal extension rate (the highest for which all the flows are considered stable and symmetric). In general, the degree of flow modification caused by viscoelasticity becomes increasingly severe with increasing polymer concentration. For uniaxial extension at $\dot\varepsilon_{nom} = 1.66~\text{s}^{-1}$ (Fig.~\ref{PAA_trans_PIV}(d)), at lower polymer concentrations of 50 and 100~ppm, the velocity profile is flattened compared with the Newtonian case, but an increasing local minimum in the centerline flow velocity develops as the PAA concentration is raised to 200~ppm and above. In planar extension at $\dot\varepsilon_{nom} = 4.16~\text{s}^{-1}$ (Fig.~\ref{PAA_trans_PIV}(e)), and low polymer concentrations of 50 and 100~ppm of PAA, the profiles are essentially Newtonian-like. For 200~ppm of polymer, the profile becomes flattened compared with the Newtonian case, and for 400~ppm a local minimum around $y=0$ becomes evident. In the case of biaxial extension at $\dot\varepsilon_{B,nom} = 13.3~\text{s}^{-1}$, flow modification by the polymer is evident at 200 and 400~ppm of PAA, where the flow velocity is reduced about the centerline and the profiles become flattened again (Fig.~\ref{PAA_trans_PIV}(f)). However, in biaxial extension, the flow profiles always remain essentially parabolic.

Velocity profiles with a local minimum in the streamwise velocity along the stretching axis have been reported a number of times in the literature studying stagnation point flows of viscoelastic fluids (e.g., Refs.~\citenum{Lyazid1980,Gardner1982,Dunlap1987,Harlen1990,Haward2010b,Haward2012c}). The reduction in flow velocity on the axis (relative to a Newtonian fluid) is associated with the localized stretching of polymers that pass near the stagnation point and are subsequently advected downstream along the outlet centerline. For polymer solutions that exhibit measurable flow-induced birefringence, this stretching results in the appearance of a characteristic `birefringent strand' localized along the stretching axis (e.g., Refs.~\citenum{Keller1985,Harlen1990,Harlen1992,Remmelgas1999,Becherer2008,Becherer2009,Haward2012c}), and indicative of high extensional stress.~\cite{Fuller} Within the strand, the fluid behaves elastically and exhibits a much higher extensional viscosity than the fluid flowing outside the strand, where the polymer is relatively unstretched, and the fluid remains Newtonian-like. The elastic strand thus acts as an internal stress boundary layer in the flow, driving velocity perturbations that resist the stretching, and thus giving rise to the modified flow profile observed.~\cite{Harlen1990}

The modification to the Newtonian flow field by the stretching of the polymer along the extensional axis reduces the true extension rate along the stretching axis, as assessed in Fig.~\ref{PAA_PIV}. Normalized profiles of the streamwise velocity component along the stretching axis are shown for the 100~ppm PAA solution over a range of nominal extension rates under uniaxial, planar, and biaxial extension in Fig.~\ref{PAA_PIV}(a), (b), and (c), respectively. Here we only consider flows that are deemed stable and symmetric. Under uniaxial extension (Fig.~\ref{PAA_PIV}(a)), the velocity profile for the 100~ppm PAA solution agrees well with the Newtonian profile for $\dot\varepsilon_{nom} \leq 0.83~\text{s}^{-1}$, but increasingly deviates from the Newtonian profile as the imposed flow rate is increased beyond $\dot\varepsilon_{nom} \approx 0.83~\text{s}^{-1}$. Under planar extension (Fig.~\ref{PAA_PIV}(b)), as $\dot\varepsilon_{nom}$ is increased, only a slight deviation from the Newtonian profile is evident even at the highest nominal extension rates tested (up to 16.7~s$^{-1}$), while for biaxial extension (Fig.~\ref{PAA_PIV}(c)), the profiles for the polymer solution remain Newtonian-like for $\dot\varepsilon_{B,nom}$ up to 53.1~s$^{-1}$ (the highest imposed value).

In Fig.~\ref{PAA_PIV}(d), (e), and (f) we plot the measured extension rate (determined from velocity profiles such as those shown in Fig.~\ref{PAA_PIV}(a), (b), and (c)) as a function of the imposed flow rate or nominal extension rate in uniaxial, planar, and biaxial extension, respectively. Here data is shown for all the tested polymer solutions and is compared against the Newtonian result (shown by the dashed grey lines). For uniaxial extension (Fig.~\ref{PAA_PIV}(d)), we report $\dot\varepsilon = \partial w/\partial z$, which is evaluated along the $z$-axis for $\abs{z} \leq 4R$ (i.e., within the range over which the flow field is optimized~\cite{Haward2023}). Similarly, in planar extension (Fig.~\ref{PAA_PIV}(e)), we report $\dot\varepsilon = \partial u/\partial x$, evaluated along the $x$-axis for $\abs{x} \leq 12W$. In biaxial extension, (Fig.~\ref{PAA_PIV}(f)), we report $\dot\varepsilon_B = \partial u'/\partial x'$, which is evaluated along the $x'$-axis for $\abs{x'} \leq R$ (the range over which axisymmetry applies such that $u'(x') \equiv u(x)$, see Fig.~\ref{NewtPIV}(f)). As shown in Fig.~\ref{PAA_PIV}(d), (e), and (f), for all three extensional flow configurations and all polymer concentrations, the polymer solutions follow the Newtonian trend for lower imposed nominal extension rates, but progressively deviate below the Newtonian trend at higher flow rates. The large polymeric stresses induced by the extensional stretching always retard the evolution of the velocity profile downstream of the stagnation point. For each polymeric fluid in each extensional flow configuration, the experimental data are well-described by a curve of the form $\dot\varepsilon = \dot\varepsilon_{nom} - A{\dot\varepsilon_{nom}}^B$ (where $A$ and $B$ are fitting constants), as shown by the respective solid lines. The fitted curves allow calculation of the true strain rate for arbitrary imposed flow conditions with the given fluid. 

In general, from Figs.~\ref{PAA_trans_PIV} and \ref{PAA_PIV} it is evident that the deviation from Newtonian-like behavior becomes more severe with increasing polymer concentration and increasing extension rate. Also, the greatest effects are observed in uniaxial extension while biaxial extension causes the mildest modification of the flow field. The effect of planar extension appears to be intermediate between uniaxial and biaxial. A similar general trend is also apparent from the onset of instability, which for a given polymer solution occurs at the lowest nominal Weissenberg number in uniaxial extension, followed by planar extension, and finally biaxial extension. This may have implications for the utility of the different flows for extensional rheometry, as will be discussed further below.

\begin{figure*}[!ht]
\begin{center}
\includegraphics[scale=0.88]{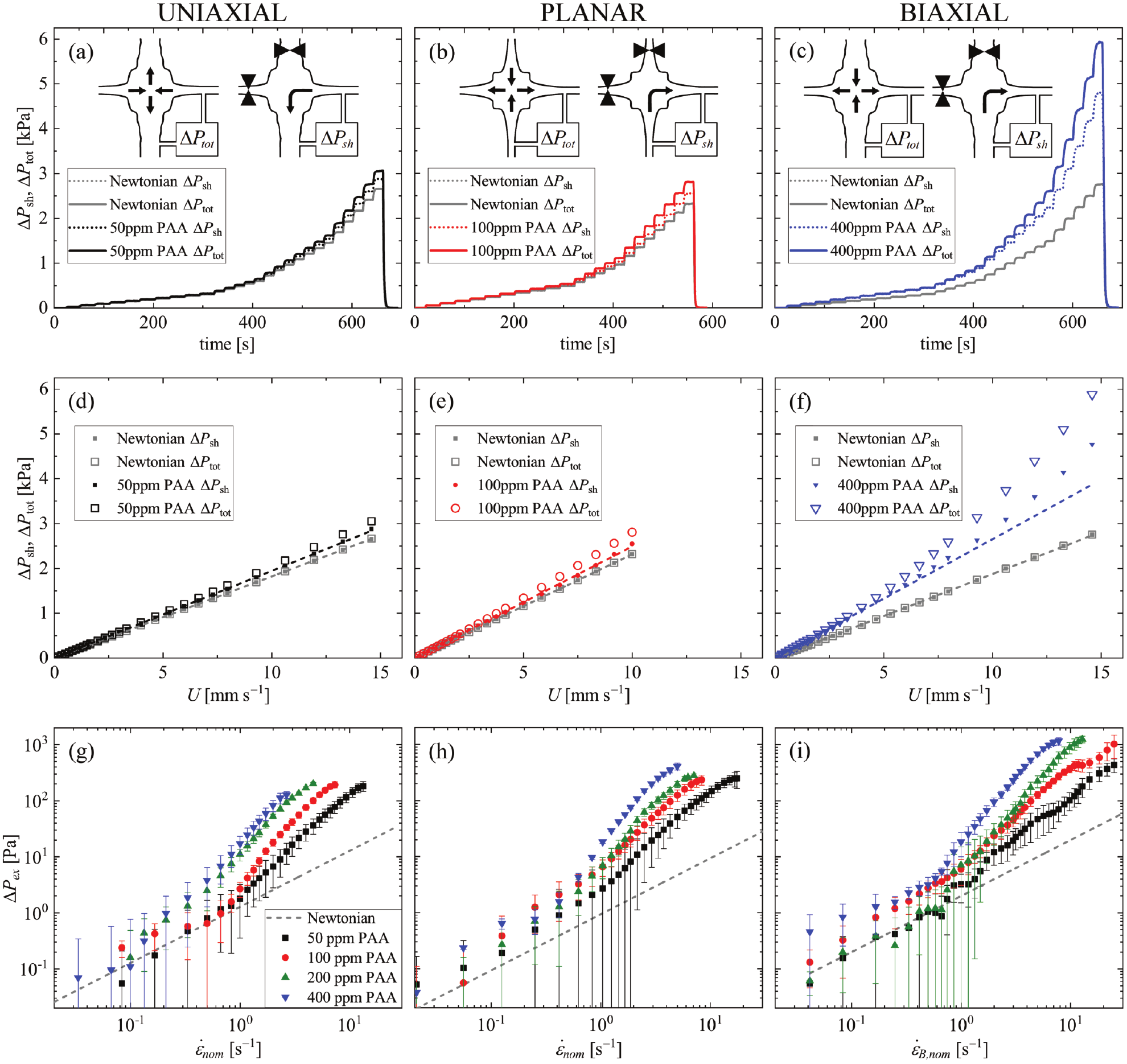}
\vspace{-0.05in}
\caption {Pressure drop measurements made with the Newtonian solvent and PAA solutions in uniaxial (left column), planar (middle column), and biaxial (right column) extensional flow. (a), (b) and (c) show representative raw measurements of the total pressure drop ($\Delta P_{tot}$) and the pressure due to shear ($\Delta P_{sh}$) versus time as the imposed flow rate $U$ is increased in steps. Inserts schematically indicate the flow configurations used for measurement of $\Delta P_{tot}$ and $\Delta P_{sh}$ in each case. The respective steady state plateau value of the pressure drop at each increment in $U$ is presented as a function of $U$ in (d), (e) and (f), where the dashed lines represent linear fits to $\Delta P_{sh}$ passing through the origin, for low $U < 2.5~\text{mm~s}^{-1}$. (g), (h), and (i) show the excess pressure drop ($\Delta P_{ex} = \Delta P_{tot}-\Delta P_{sh}$) for all of the tested fluids as a function of the nominal strain rate in uniaxial, planar, and biaxial extension, respectively. Dashed grey lines are linear fits through the data for the Newtonian fluid, with constants of proportionality $\approx1.3$~Pa~s, $\approx0.9$~Pa~s, and $\approx2.0$~Pa~s in parts (g), (h), and (i), respectively. Error bars on $\Delta P_{ex}$ for the polymer solutions represent the standard deviation over at least five repeated measurements.
} 
\label{pressure}
\end{center}
\end{figure*}

\subsection{Pressure drop}

In Fig.~\ref{pressure}(a), (b), and (c), we illustrate raw measurements of the pressure drop in uniaxial, planar, and biaxial extension (respectively) using a few of the polymeric fluids and also the Newtonian solvent. As described in Sec.~\ref{press} (and illustrated schematically by the respective inserts to Fig.~\ref{pressure}(a), (b), and (c)), for each fluid and each flow configuration, the pressure drop is measured once with the device in full operation mode to obtain the total pressure drop $\Delta P_{tot}$, and once with half of the inlet channels and half of the outlet channels disabled in order to quantify the contribution of shear $\Delta P_{sh}$. The measurements are made by programming the syringe pumps to increment the average flow velocity through the device $U$ in a stepwise fashion, with sufficient time at each step for the pressure to rise and stabilize to a steady plateau value. Subsequently, the average plateau pressure drop is measured at each step in flow rate in order to obtain curves such as those shown in Fig.~\ref{pressure}(d), (e), and (f), which result from the raw pressure traces shown in Fig.~\ref{pressure}(a), (b), and (c), respectively. Note that in each flow configuration, for the Newtonian fluid $\Delta P_{tot} \approx \Delta P_{sh}$, with $\Delta P_{sh} \propto U$ (as indicated by the dashed grey lines). For lower concentration polymer solutions, $\Delta P_{sh} \propto U$ (as indicated by the dashed black and red lines in Fig.~\ref{pressure}(d) and (e), repsectively), although at higher polymer concentrations, $\Delta P_{sh}$ may increase superlinearly at higher imposed flow rates (as shown by the deviation of the experimental data points from the dashed blue line in Fig.~\ref{pressure}(f)). Most notably, for the polymer solutions at low average flow velocities $\Delta P_{tot} \approx \Delta P_{sh}$ (as we also observe for the Newtonian fluid), but beyond a certain value of $U$, the two curves diverge and $\Delta P_{tot}$ rises clearly above $\Delta P_{sh}$, leading to a significant and clearly measurable excess pressure drop $\Delta P_{ex} = \Delta P_{tot} - \Delta P_{sh}$.

In Fig.~\ref{pressure}(g), (h), and (i), we present $\Delta P_{ex}$ as a function of the nominal extension rate for all of the tested fluids under uniaxial, planar, and biaxial extension (respectively). Here, error bars represent the standard deviation over a minimum of five repeated measurements. Scatter and uncertainty in the data for the Newtonian solvent fluid is significant, so for better clarity the data in each plot is represented by a linear fit (dashed grey line). In general, at low nominal deformation rates the data from the polymer solutions follow a roughly linear trend (similar to the Newtonian fluid), but depart from that trend as the extension rate increases, turning upwards and tending towards eventual plateau values. Note that, in several cases, the measured pressure difference ($\Delta P_{tot}$ and/or $\Delta P_{sh}$) exhibits fluctuations at higher extension rates (even though the flow field may be deemed steady and symmetric, see Sec.~\ref{flowfield}). For this reason, the maximum extension rates at which data is curtailed in Fig.~\ref{PAA_PIV}(d), (e), and (f) and in Fig.~\ref{pressure}(g), (h), and (i) (respectively) may not always precisely match, and in some cases there is an increase in the reported error bars in $\Delta P_{ex}$ at the highest rates tested.

\subsection{Extensional rheometry}
\label{SecetaE}

With the required experimental data, i.e., true measured extension rates, and excess pressure drops in hand (see Figs.~\ref{PAA_PIV} and~\ref{pressure}, respectively), we turn to simple theory in order to understand how to obtain a robust estimate of the extensional viscosity from our measurements. 

Considering for simplicity the 2D flow through the OSCER geometry, a macroscopic power balance leads to the following approximate expression (see Appendix~\ref{appendixA} for details): 

\begin{equation}
2\Delta P_{ex} Q \approx (\sigma_{xx}-\sigma_{yy})\dot\varepsilon\mathcal{V}_P
,
\label{OSCERDtau}
\end{equation}
which allows estimation of the extensional viscosity:
\begin{equation}
\eta_{P} \approx 2\Delta P_{ex} Q/ \dot\varepsilon^2\mathcal{V}_P
,
\label{eta_p}
\end{equation}
where $\mathcal{V}_P$ is an appropriate volume of fluid within the device over which $\sigma_{xx}-\sigma_{yy}$ can be considered `homogeneous' for averaging purposes. 

For a Newtonian fluid, or for a viscoelastic fluid flowing at $\text{Wi} \ll 0.5$, $\mathcal{V}_P$ might be expected to roughly equate with the volume of the optimized region of the OSCER geometry, $V_{OSC} =480W^2H~(=4.8 \times10^{-9}~\text{m}^3$ for the specific device being used here with $H=1$~mm and $W=0.1$~mm, Sec.~\ref{geom}).

Since for a Newtonian fluid in planar extension the Trouton ratio is known to be $\text{Tr}=(\sigma_{xx}-\sigma_{yy})/\dot\varepsilon \eta = 4$, Eq.~\ref{OSCERDtau} can be rewritten and rearranged to give:

\begin{equation}
\mathcal{V}_{P,Newt} = 2 \Delta P_{ex} Q / 4\dot\varepsilon^2 \eta
.
\label{V_Newt}
\end{equation}
We know that for Newtonian flow in the OSCER device, $\dot\varepsilon \approx \dot\varepsilon_{nom} = 0.1U/W = 0.1Q/4W^2H$. Furthermore, from the fit to the Newtonian data shown in Fig.~\ref{pressure}(h), we know that $\Delta P_{ex} \approx 0.9 \dot\varepsilon$. Hence, Eq.~\ref{V_Newt} can be evaluated to give a unique value $\mathcal{V}_{P,Newt}\approx126W^2H$  (or $\approx 0.25V_{OSC}$). We should indeed anticipate that $\mathcal{V}_{P,Newt} < V_{OSC}$ since the extensional kinematics are not entirely homogeneous over the whole OSCER geometry due to the shear induced at the channel walls.~\cite{Haward2012c,Haward2016c} In fact, a volume of magnitude $0.25V_{OSC}$ corresponds well to the volume of the OSCER geometry over which the local extension rate is within 10\% of $\dot\varepsilon_{nom}$, i.e., the region where the extensional kinematics are almost homogeneous.

As discussed in Sec.~\ref{flowfield} and elsewhere, for a viscoelastic fluid in an extensional flow at $\text{Wi} > 0.5$, a localized elastic `birefringent strand' develops along the stretching axis within which $\Delta \sigma$ becomes dominant.~\cite{Harlen1990,Harlen1992,Becherer2008,Becherer2009,Haward2012c} Accordingly, we expect that for $\text{Wi} > 0.5$, the relevant volume $\mathcal{V}_P$ to use in Eqs.~\ref{OSCERDtau} or \ref{eta_p}, would be that of the birefringent strand.

In principle, in certain cases, it may be possible to directly measure the dimensions of the birefringent elastic strand in order to determine its volume experimentally as a function of the Weissenberg number. However, this is not always practical or even possible; for instance in the present case, the fluids being used are too weakly birefringent to make the required optical measurements. For this reason, we seek a simple and pragmatic approach to estimate the volume of the elastic strand $\mathcal{V}_{P,strand}$, which may be used more generally in Eqs.~\ref{OSCERDtau} and \ref{eta_p} when $\text{Wi} \geq 0.5$.

For the FENE-P model in planar extension, an approximate scaling relation for the dimensionless half-width $w_{strand}^*$ of the sheet-like birefringent strand in terms of $\text{Wi}$ and the polymer extensibility $L$ has been presented by Becherer et al.~\cite{Becherer2008} The scaling has been shown to adequately describe measurements of the birefringent strands that develop in the OSCER device for $\text{Wi} \geq 0.5$.~\cite{Haward2012c} In the asymptotic limit of high $\text{Wi}$, the dimensionless strand half-width scales as $w_{strand}^* \sim 1/L$.~\cite{Crowley1976,Rallison1988,Harlen1992,Renardy2006,Becherer2008} A dimensional strand half-width can be computed as $w_{strand} \approx l_{opt}/L$, where $l_{opt}=15W$ is the lengthscale over which the flow field in the OSCER device is optimized and the flow is purely extensional.

Accordingly, we approximate the asymptotic volume of the sheet-like birefringent strand as $\mathcal{V}_{P,strand} \approx 1800 W^2H/L$ (strand length $l_{strand}=30W$, width $2w_{strand} = 30W/L$, height $h_{strand} = 2H$), which for $L=143$ (Sec.~\ref{fluids}) yields $\mathcal{V}_{P,strand} \approx \mathcal{V}_{P,Newt}/10$. We propose the following simple piecewise approximation to the volume $\mathcal{V}_P$ as a function of $\text{Wi}$ for planar extensional flow of dilute solutions of flexible polymers in the OSCER device:


\begin{equation}
\mathcal{V}_P=
\begin{cases}
\mathcal{V}_{P,Newt} \approx 126W^2H & \text{:} \quad \text{Wi}<0.5  \\
\mathcal{V}_{P,strand} \approx 1800W^2H/L  & \text{:}  \quad \text{Wi}\geq0.5 \\
\end{cases}
.
\label{Vp}
\end{equation}

By following similar arguments, we arrive at the following equations to evaluate $\eta_E$ in the case of uniaxial extensional flow in the OUBER device:

\begin{equation}
2\Delta P_{ex} Q \approx (\sigma_{zz}-\sigma_{xx})\dot\varepsilon\mathcal{V}_E
,
\label{OUBER_uni_Dtau}
\end{equation}
i.e.:
\begin{equation}
\eta_{E} \approx 2\Delta P_{ex} Q/ \dot\varepsilon^2\mathcal{V}_E
,
\label{eta_E}
\end{equation}
where 
\begin{equation}
\mathcal{V}_E=
\begin{cases}
\mathcal{V}_{E,Newt} \approx 47R^3  & \text{:} \quad \text{Wi}<0.5  \\
\mathcal{V}_{E,strand} \approx 250\uppi R^3/L  & \text{:}  \quad \text{Wi}\geq0.5 \\
\end{cases}
.
\label{VE}
\end{equation}
Here, $\mathcal{V}_{E,Newt} \approx 0.31 V_{OUB}$, where $V_{OUB} =154R^3$ is the volume of the optimized region of the OUBER device. For the particular OUBER device being used in this study, with $R=0.4$~mm (Sec.~\ref{geom}), $V_{OUB} \approx 9.86 \times 10^{-9}~\text{m}^3$. For uniaxial extension, Harlen et al (1992)~\cite{Harlen1992} have shown for the FENE-CR model that the asymptotic radius of the birefringent strand at high $\text{Wi}$ scales as $1/\sqrt{L}$, which is consistent with experimental measurements made in classical opposed-jets apparatus,~\cite{Muller1988,Cathey1990} as well as in a 6-arm cross-slot device.~\cite{Haward2019b} In Eq.~\ref{VE}, $\mathcal{V}_{E,strand}$ represents the volume of a columnar birefringent strand of asymptotic diameter $10R/\sqrt{L}$ and length $10R$.

\begin{figure*}[ht]
\begin{center}
\includegraphics[scale=0.59]{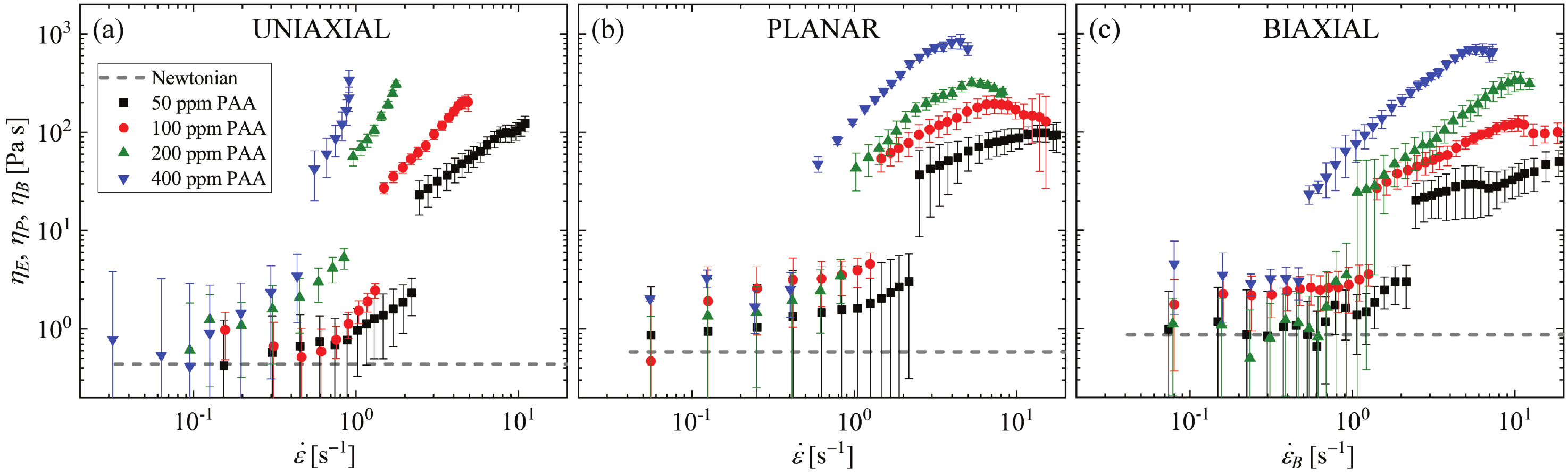}
\caption {Extensional viscosity as a function of the extensional strain rate for the Newtonian solvent and the dilute PAA solutions, determined from excess pressure drop measurements made in (a) uniaxial, (b) planar, and (c) biaxial elongational flow. 
} 
\label{etaE}
\end{center}
\end{figure*}

For biaxial extensional flow in the OUBER device, we obtain:

\begin{equation}
\Delta P_{ex} Q \approx (\sigma_{xx}-\sigma_{zz})\dot\varepsilon_B\mathcal{V}_B
,
\label{OUBER_bi_Dtau}
\end{equation}
i.e.:
\begin{equation}
\eta_{B} \approx \Delta P_{ex} Q/ \dot\varepsilon_B^2\mathcal{V}_B
,
\label{eta_B}
\end{equation}
where 

\begin{equation}
\mathcal{V}_B=
\begin{cases}
\mathcal{V}_{B,Newt} \approx 37.5R^3 & \text{:} \quad \text{Wi}<0.5  \\
\mathcal{V}_{B,strand} \approx 250\uppi R^3/L  & \text{:}  \quad \text{Wi}\geq0.5 \\
\end{cases}
,
\label{VB}
\end{equation}
and in this case $\mathcal{V}_{B,Newt} \approx 0.24 V_{OUB}$. For biaxial stagnation point extension, the thickness of the disk-like birefringent ``strand'' region (e.g., Refs.~\citenum{Frank1971,Backus2002}) that forms over the $z=0$ plane has not been well characterized in the literature. Additionally, since there is a disagreement between constitutive models regarding the response of polymeric solutions to biaxial extension, we do not wish to rely on the prediction of a specific (e.g., FENE-type) model to describe the dimension of the resulting birefringent region. Limited experimental data obtained from a dilute solution of near-monodisperse atactic polystyrene in a 6-arm cross-slot device over a range of $\text{Wi}$ in both uniaxial and biaxial elongation is available in Ref.~\citenum{Haward2019b}. Assuming that the asymptotic strand radius in uniaxial extension scales as $1/\sqrt{L}$ (as established above), the data available in Ref.~\citenum{Haward2019b} indicates the asymptotic thickness of the birefringent region in biaxial extension to scale as $\sim 1/L$,~\cite{Haward2019b} similar to the result for planar elongation.~\cite{Becherer2008,Haward2012c} Accordingly, in Eq.~\ref{VB}, $\mathcal{V}_{B,strand}$ represents the volume of a birefringent disk of asymptotic thickness $10R/L$ and diameter $10R$.

Note that the computation of $\mathcal{V}_{P,Newt}$, $\mathcal{V}_{E,Newt}$, and $\mathcal{V}_{B,Newt}$ using the excess pressure drop measured for the Newtonian fluid serves as a Newtonian calibration of the respective flow, ensuring the correct value of $\text{Tr}$ will be obtained for the Newtonian fluid when those volumes are used to compute the extensional viscosity from Eqs.~\ref{eta_p}, \ref{eta_E}, and \ref{eta_B}, respectively.

The volumes computed for the Newtonian fluid and for the birefringent strand regions (given in Eqs.~\ref{Vp}, \ref{VE}, and \ref{VB}), should be valid for $\text{Wi}\rightarrow0$ and $\text{Wi}\rightarrow \infty$, respectively. The step change in $\mathcal{V}_P$, $\mathcal{V}_E$, and $\mathcal{V}_B$ at $\text{Wi}=0.5$ is clearly unphysical, however at present the functional form that the volume should take across this transition between Newtonian-like and viscoelastic behavior is unclear. The question over this is further complicated if we are to consider a $\text{Wi}$-dependent strand volume, which vanishes for $\text{Wi}\leq0.5$,~\cite{Becherer2008} suggesting a possibly nonmonotonic variation of the volume with $\text{Wi}$. Despite this shortcoming, we consider the formulation described above to be an advance on earlier estimates of the extensional viscosity from pressure drop measurements. For instance, in the cross-slot and OSCER devices, the rather coarse approximation $\eta_P \approx \Delta P_{ex}/\dot\varepsilon$ was generally used (e.g., Ref.~\citenum{Haward2012c}), although some prior attempts have also been made to account for the dimensions of the birefringent strand.~\cite{Haward2010,Haward2010b} Given the experimentally-established linear relation between $Q$ and $\dot\varepsilon$ (at least for the Newtonian fluid) it can be seen that our new approximation to the planar extensional viscosity measured in the OSCER device (Eq.~\ref{eta_p}) can be written $\eta_P \approx (\Delta P_{ex}/\dot\varepsilon) \times F$, where $F = 8W^2H/0.1\mathcal{V}_P$ is a dimensionless correction factor essentially consisting of a ratio of geometric parameters. An analogy can be drawn with the determination of the shear viscosity from the experimentally-measured pressure drop along a pipe or channel of arbitrary cross-section, where the pressure drop must be scaled by the ratio of the hydraulic diameter to the length of the conduit.~\cite{Walters}

The extensional viscosities $\eta_E(\dot\varepsilon)$, $\eta_P(\dot\varepsilon)$, and $\eta_B(\dot\varepsilon_B)$, computed as described above, are shown for each of the experimental test fluids in Fig.~\ref{etaE}(a), (b), and (c), respectively. In each plot, the Newtonian result (dashed gray line) is computed using the respective fit to the excess pressure drop data shown in Fig.~\ref{pressure}(g,h,i), resulting in a constant value for the extensional viscosity equal to $3\eta_s$, $4\eta_s$, and $6\eta_s$ in uniaxial, planar and biaxial extension, respectively. At low extension rates, the results obtained for the dilute polymer solutions generally approach a constant value close to (or slightly higher than) that of the Newtonian solvent, as expected given the slightly higher shear viscosities of the polymer solutions (Table~\ref{tab1}). With increasing extension rate, each of the polymeric fluids undergo a gradual increase in the extensional viscosity, before an abrupt jump takes place at a specific extension rate (corresponding to $\text{Wi}=0.5$) that reduces with the polymer concentration (due to the increasing relaxation time, see Table~\ref{tab1}). Subsequently, for further increasing extension rate, there is a general trend for the extensional viscosity to gradually increase towards an apparent plateau. 

\begin{figure*}[ht]
\begin{center}
\includegraphics[scale=0.75]{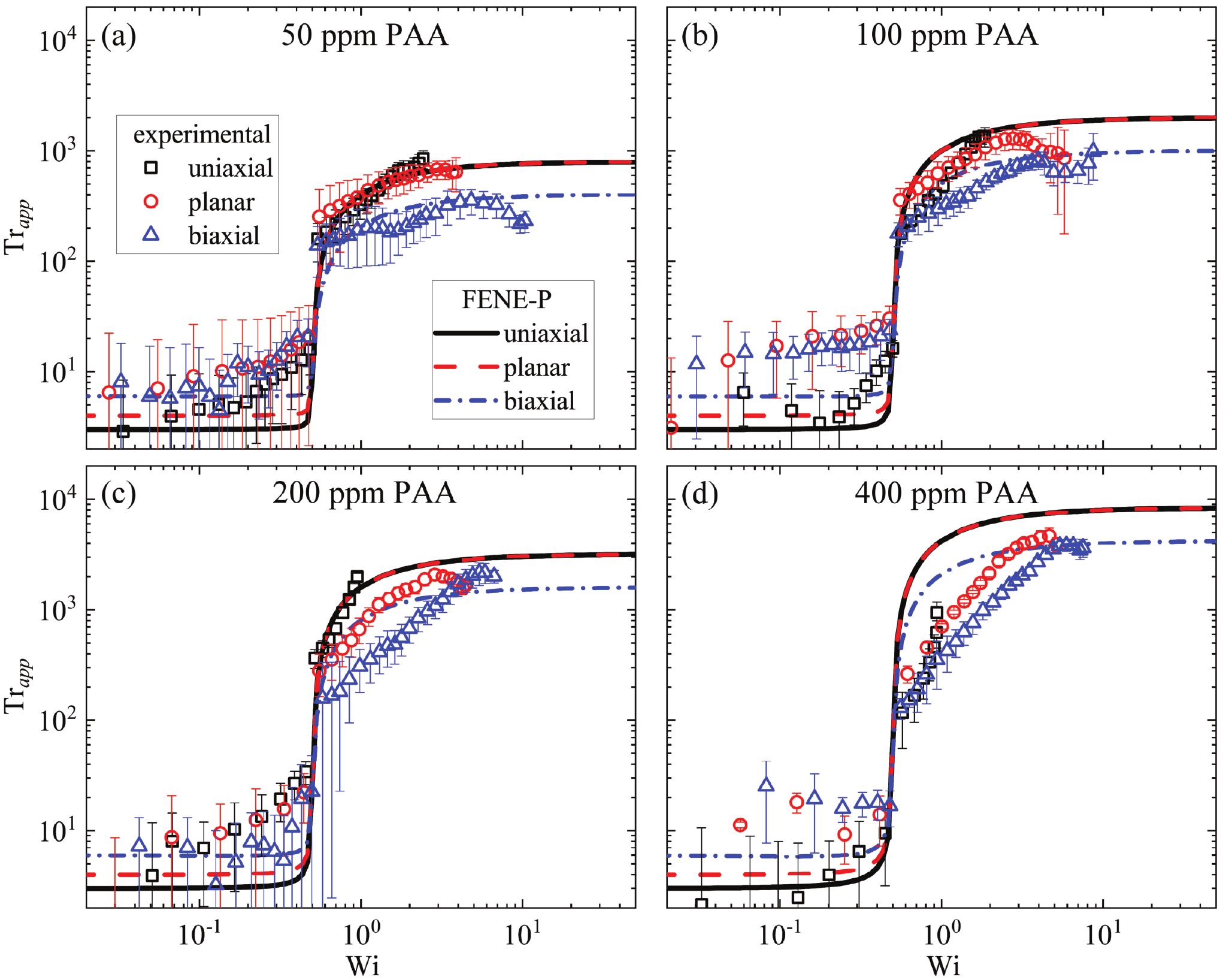}
\caption {Apparent Trouton ratio $\text{Tr}_{app}$ as a function of the Weissenberg number $\text{Wi}$ for (a) 50 ppm PAA, (b) 100 ppm PAA, (c) 200 ppm PAA, and (d) 400 ppm PAA in uniaxial, planar and biaxial elongational flow. Data points are experimentally determined from pressure loss measurements. Lines are computed from the FENE-P model with the solvent-to-total viscosity ratio $\beta$ matched to the respective fluid (given in Table~\ref{tab1}), and the extensibility parameter (or stretch ratio) $L=143$ (Sec.~\ref{fluids}).
} 
\label{Tr_vs_Wi}
\end{center}
\end{figure*}

Similarities and differences between the responses of the fluids to the different modes of extensional flow are made more obvious by viewing their apparent Trouton ratio as a function of the Weissenberg number in Fig.~\ref{Tr_vs_Wi}. Here, we also plot the response predicted by the FENE-P model under homogeneous uniaxial, planar and biaxial elongation conditions, where the model parameters $\beta$ and $L^2$ are matched to those of the fluids (solvent-to-total viscosity ratio $\beta$ given in Table~\ref{tab1}, and extensibility $L=143$, as computed in Sec.~\ref{fluids}). In general, within experimental uncertainty, for all polymer solutions at low $\text{Wi}$, $\text{Tr}_{app}$ approaches the expected (i.e., Newtonian) limiting value. Also, consistent with the model prediction, in all cases $\text{Wi}=0.5$ marks the point of an abrupt increase in $\text{Tr}_{app}$. For the most dilute 50~ppm PAA solution (Fig.~\ref{Tr_vs_Wi}(a)), for $\text{Wi}>0.5$ the experimental data obtained from uniaxial, planar, and biaxial extension closely follow the respective FENE-P prediction towards the high-$\text{Wi}$ plateau, where $\text{Tr}_{app}(\text{uniaxial}) = \text{Tr}_{app}(\text{planar}) =2\times \text{Tr}_{app}(\text{biaxial})$. As the PAA concentration is increased through Fig.~\ref{Tr_vs_Wi}(b), (c), and (d), the agreement with the FENE-P model prediction becomes less convincing, with an increasingly gradual approach of the experimental data towards the eventual plateau in the extensional viscosty and a less distinct difference between the response in biaxial extension from that in uniaxial and planar extension. These changes with polymer concentration may be because the polymer solutions (although dilute with $c/c^* \leq 0.1$) can not all be considered ``ultradilute'', and at higher polymer concentrations intermolecular interactions may play an increasingly important role when the molecules become stretched by the flow.~\cite{Dunlap1987,Harrison1998,Clasen2006,Stoltz2006} Experimental data from uniaxial extension~\cite{Clasen2006} and molecular dynamics simulations in planar extension~\cite{Stoltz2006} suggest that the ultradilute limit, for which interchain interactions are negligible even at high polymer extensions, is approached as the polymer concentration is decreased towards $c/c^* \approx 0.01$, similar to the concentration regime of our 50~ppm PAA solution. Notably, at the higher polymer concentrations tested  (Fig.~\ref{Tr_vs_Wi}(c,d)), we are unable to see a convincing high-Weissenberg number plateau in $\text{Tr}_{app}$ for uniaxial extensional flow. In these cases, the onset of elastic flow instability curtails the measurement before a plateau is reached. In fact, the flow modification is so severe in these cases (see Fig.~\ref{PAA_PIV}(d)) that $\dot\varepsilon$ almost ceases to increase with the imposed flow velocity. Since the excess pressure drop across the device continues to increase with the imposed flow velocity (Fig.~\ref{pressure}(g)), this causes an apparent upturn in $\eta_E$ and $\text{Tr}_{app}$ to an asymptote as $\text{Wi}\rightarrow1$, before the flow field breaks symmetry.

\section{Discussion and Conclusions}
\label{SumCon}

%

In this work we have used the new OUBER device (developed in Part I of this paper~\cite{Haward2023}), and also the pre-existing OSCER device,~\cite{Haward2012c} to perform the first experimental comparison of the extensional rheology of dilute mobile polymer solutions in planar, uniaxial and biaxial extensional flow. In each case the extensional viscosity is assessed using common methods: micro-particle image velocimetry is used to quantify the relevant extensional strain rate along the stretching axis (or axes) and excess pressure drop measurements are used to estimate the respective tensile stress difference as a function of the extension rate. The estimate of the tensile stress difference is based on a new analysis of the macroscopic power balance for each extensional flow configuration. In each case, the Reynolds number of the flow is maintained sufficiently low for inertial contributions to the pressure drop to be ignored. 

Several differences are observed between the responses of the polymer solutions in the various extensional flow configurations. Specifically, for a given nominal extension rate, the flow field is most severely modified (compared to that of a Newtonian fluid) in uniaxial extension. By contrast, in biaxial extensional flow of the polymer solutions the kinematics remain essentially Newtonian-like even at much higher nominal extension rates. Planar extensional deformations of the polymer solutions have an intermediate effect, showing more significant flow modification than in biaxial extension, but being less severe than in uniaxial extension. Stability constraints follow a similar trend: for a given polymer solution, uniaxial flow destabilizes and becomes asymmetric at the lowest extension rate, while biaxial flow remains stable to much higher extension rates, with planar flow being intermediate.

Our estimates of the extensional viscosities and apparent Trouton ratios of the polymer solutions, based on our new analysis method, are broadly consistent with the predictions of the FENE-P constitutive model. Within experimental error, the data approach the expected limiting values at low extension rates or Weissenberg numbers, and all of the polymeric test solutions exhibit an increase in the extensional viscosity (or $\text{Tr}_{app}$) at $\text{Wi}=0.5$. For $\text{Wi}>0.5$, the extensional viscosities of the polymeric fluids generally approach towards high-$\text{Wi}$ plateau values. For our most dilute 50~ppm polymer solution (for which $c/c^* \approx0.01$ and which can be considered ``ultradilute''), the high-$\text{Wi}$ plateau values of the extensional viscosity agree very well with the prediction of the FENE-P model, for which $\eta_E = \eta_P = 2\eta_B$. However, this agreement progressively deteriorates with increasing polymer concentration. This is likely because, although the polymer chains are dilute and non-interacting under quiescent conditions (with $c/c^* \leq 0.1$), interchain interactions become increasingly important at higher concentrations as the molecules unravel in the extensional flow. 

From a practical point of view an important consideration is the early onset of instability in the uniaxial extensional flow. This can cause difficulty in reaching the high-$\text{Wi}$ plateau of the extensional viscosity, and therefore limits the utility of the device for measurement of $\eta_E$. The greater relative stability of planar and biaxial extensional flows allow measurements to be made to much higher extension rates (or larger limiting Weissenberg numbers), and for plateau values of $\eta_P$ and $\eta_B$ to be found more convincingly. On the other hand, it will be of fundamental interest to understand the three-dimensional form of the symmetry-breaking flow instability that occurs in uniaxial extension, which is not readily ascertained from the 2D flow velocimetry performed in the present work (see Fig.~\ref{100ppmPIV}(g)). It will also be important to better understand the physical reason for why uniaxial extension is the most prone to the onset of elastic instability. We speculate at present that this is related to the greater thickness of the birefringent strand (radius $\sim 1/\sqrt{L}$ in uniaxial extension, but half-width $\sim 1/L$ in planar and biaxial extension). We think this is likely to be the reason why the flow modification along the stretching axis is most severe in the case of uniaxial extension, and that this probably contributes to the onset of instability, too. 

We reiterate that our estimates of the extensional viscosities $\eta_E,~\eta_P,$ and $\eta_B$ are just that (i.e., \emph{estimates}), as will necessarily always be the case since generating a spatially homogeneous extensional flow throughout the whole of the rheometric device is practically impossible. However, we have for the first time designed and fabricated microfluidic devices that generate reasonably homogeneous approximations to uniaxial, planar and biaxial extension over spatial regions much larger than the characteristic lengthscale of the geometry, and which also permit comparable assessments to be made of the tensile stress difference as a function of the imposed extension rate, all at low levels of fluid inertia. We believe that our new approach to estimating the tensile stress difference from the excess pressure drop, based on an approximate solution to the macroscopic power balance (see Sec.~\ref{SecetaE}) represents a significant advance on prior analyses in similar such devices. Nevertheless, there remains significant scope for further improvement. Specifically, at the transition between Newtonian-like and viscoelastic behavior at $\text{Wi}=0.5$, the abrupt step down in the volume used to compute the extensional viscosity in Eqs.~\ref{eta_p}, \ref{eta_E}, and \ref{eta_B} is unphysical. Clearly this transition should be smooth, but at present it is unclear how it should be described mathematically. Numerical simulations may provide insight to this computational rheology problem, although it is possible that polydispersity of the polymer molecular weight also contributes to the form of this transition region, which may be confounding. Furthermore, the estimation of the volume of the birefringent strand for $\text{Wi}>0.5$ should strictly depend on $\text{Wi}$, which further complicates the analysis. An analytical solution (based on the FENE-P model) for the width of the birefringent strand as a function of $\text{Wi}$ is available for planar extension,~\cite{Becherer2008} but the corresponding elastic boundary layer analysis needs to be solved (and confirmed experimentally) for uniaxial and biaxial extension. In our ongoing work, we intend to focus our research efforts towards addressing these issues.

\begin{acknowledgments}
S.J.H, S.V., and A.Q.S. gratefully acknowledge the support of the Okinawa Institute of Science and Technology Graduate University (OIST) with subsidy funding from the Cabinet Office, Government of Japan, along with funding from the Japan Society for the Promotion of Science (JSPS, Grant Nos. 21K03884 and 22K14184). MAA acknowledges the support by LA/P/0045/2020 (ALiCE), UIDB/00532/2020 and UIDP/00532/2020 (CEFT), funded by national funds through FCT/MCTES (PIDDAC). We are indebted to Prof. R. J.  Poole for insightful discussions.
\end{acknowledgments}

\section*{Conflict of Interest Statement}

The authors have no conflicts to disclose.

\section*{Data Availability Statement}

The data that support the findings of this study are available from the corresponding author upon reasonable request.

\appendix

\section{Origin of $\Delta P_{ex}$ in the OSCER device}
\label{appendixA}

We first consider 2D flow in the OSCER device with flow injected at volumetric rate $Q$ through two opposing inlet channels and withdrawn at volumetric rate $Q$ through two opposing outlet channels (i.e., normal operation mode) to generate the stagnation point extensional flow field. Assuming that the flow is fully developed at the inflow and outflow boundaries, the total pressure drop $\Delta P_{tot}$ measured across an inlet and an outlet of the device must obey the following macroscopic power balance equation:

\begin{equation}
\Delta P_{tot} Q  = \int\limits_{V/2}\biggl(\sigma_{xx}\frac{\partial u}{\partial x} + \sigma_{xy}\Bigl(\frac{\partial u}{\partial y} + \frac{\partial v}{\partial x}\Bigr) + \sigma_{yy}\frac{\partial v}{\partial y}\biggr)dV                      
,
\label{A1}
\end{equation}
which by application of the continuity equation can be written as:

\begin{equation}
\begin{split}
\Delta P_{tot} Q & =   \int\limits_{V/2} (\sigma_{xx}-\sigma_{yy})\frac{\partial u}{\partial x}dV + \int\limits_{V/2} \sigma_{xy}\Bigl(\frac{\partial u}{\partial y} + \frac{\partial v}{\partial x}\Bigr)dV
.
\end{split}
\label{A2}
\end{equation}


We split the total volume $V$ into two parts: $V_{OSC}$, the extensional flow-dominated volume of the optimized region of the OSCER device, and $V-V_{OSC}$, which comprises the remaining regions between the upstream and downstream pressure taps (i.e., shear-dominated regions in the inlets, outlets, and connecting tubing). Thus Eq.~\ref{A2} can be rewritten:

\begin{equation}
\begin{split}
\Delta P_{tot} Q & = \int\limits_{V_{OSC}/2}(\sigma_{xx}-\sigma_{yy})\frac{\partial u}{\partial x}dV + \int\limits_{(V-V_{OSC})/2}(\sigma_{xx}-\sigma_{yy})\frac{\partial u}{\partial x}dV \\
                        & + \int\limits_{V_{OSC}/2}\sigma_{xy}\Bigl(\frac{\partial u}{\partial y} + \frac{\partial v}{\partial x}\Bigr)dV + \int\limits_{(V-V_{OSC})/2}\sigma_{xy}\Bigl(\frac{\partial u}{\partial y} + \frac{\partial v}{\partial x}\Bigr)dV 
.
\end{split}
\label{A3}
\end{equation}

Since the flow is shear-dominated in the volume $V-V_{OSC}$, and extension-dominated in the volume $V_{OSC}$ the second and third terms in Eq.~\ref{A3} can be neglected, leaving:

\begin{equation}
\Delta P_{tot} Q  \approx \int\limits_{V_{OSC}/2}(\sigma_{xx}-\sigma_{yy})\frac{\partial u}{\partial x}dV  + \int\limits_{(V-V_{OSC})/2}\sigma_{xy}\Bigl(\frac{\partial u}{\partial y} + \frac{\partial v}{\partial x}\Bigr)dV 
.
\label{A4}
\end{equation}

Similar to the derivation of Eq.~\ref{A3}, the pressure drop measured when one inlet and one outlet of the OSCER device is disabled ($\Delta P_{sh}$) must obey the following:

\begin{equation}
\begin{split}
\Delta P_{sh} Q & = \int\limits_{V_{OSC}}(\sigma_{xx}-\sigma_{yy})\frac{\partial u}{\partial x}dV + \int\limits_{(V-V_{OSC})/2}(\sigma_{xx}-\sigma_{yy})\frac{\partial u}{\partial x}dV \\
                        & + \int\limits_{V_{OSC}}\sigma_{xy}\Bigl(\frac{\partial u}{\partial y} + \frac{\partial v}{\partial x}\Bigr)dV + \int\limits_{(V-V_{OSC})/2}\sigma_{xy}\Bigl(\frac{\partial u}{\partial y} + \frac{\partial v}{\partial x}\Bigr)dV 
.
\end{split}
\label{A5}
\end{equation}

Since, in this case, the extensional component in the flow is absent in the entire volume, the first and second terms in Eq.~\ref{A5} can be neglected. We also neglect the third term since the shear rate in $V_{OSC}$ is much smaller than that in the inlet and outlet regions ($V-V_{OSC}$). Thus:

\begin{equation}
\Delta P_{sh} Q \approx \int\limits_{(V-V_{OSC})/2}\sigma_{xy}\Bigl(\frac{\partial u}{\partial y} + \frac{\partial v}{\partial x}\Bigr)dV 
.
\label{A6}
\end{equation}

Subtracting Eq.~\ref{A6} from Eq.~\ref{A4} to solve for the excess pressure drop $\Delta P_{ex} = \Delta P_{tot} - \Delta P_{sh}$, we obtain:

\begin{equation}
\Delta P_{ex} Q \approx \int\limits_{V_{OSC}/2}(\sigma_{xx}-\sigma_{yy})\frac{\partial u}{\partial x}dV
.
\label{A7}
\end{equation}

The integrated volume $V_{OSC}/2$ can be further subdivided if the tensile stress difference $\sigma_{xx} - \sigma_{yy}$ becomes dominant in a specific region, such as within the birefringent strand when $\text{Wi} > 0.5$.~\cite{Haward2012c} Thus:

\begin{equation}
\Delta P_{ex} Q \approx \int\limits_{\mathcal{V}_P/2}(\sigma_{xx}-\sigma_{yy})\frac{\partial u}{\partial x}dV + \int\limits_{(V_{OSC}-\mathcal{V}_P)/2}(\sigma_{xx}-\sigma_{yy})\frac{\partial u}{\partial x}dV
,
\label{A8}
\end{equation}
where $\mathcal{V}_P$ represents the relevant volume of fluid over which the tensile stress difference should be averaged. Accordingly, the second term in Eq.~\ref{A8} can be dropped. Assuming homogeneous extension within the volume $\mathcal{V}_P$ (e.g., the strand,~\cite{Becherer2008}) we obtain:

\begin{equation}
2\Delta P_{ex} Q \approx (\sigma_{xx}-\sigma_{yy})\dot\varepsilon\mathcal{V}_P
.
\label{A9}
\end{equation}

\section*{References}


%

\end{document}